\RequirePackage{fix-cm}
\documentclass[smallextended]{svjour3}       
\smartqed  
\usepackage{graphicx}
\usepackage{mathptmx}      
\usepackage{latexsym}
\usepackage{amssymb}
\usepackage{amsmath}
\usepackage{color}
\usepackage{bm}
\newcommand{\beq}{\begin{equation}}
\newcommand{\eeq}{\end{equation}}
\newcommand{\nn}{\nonumber}
\newcommand{\om}{\omega}
\newcommand{\Om}{\Omega}
\newcommand{\g}{\gamma}
\newcommand{\gt}{\tilde{\g}}

\newcommand{\p}{\psi}
\newcommand{\s}{\sigma}
\newcommand{\D}{\Delta}
\newcommand{\del}{\delta}

\newcommand{\ra}{\rightarrow}
\newcommand{\ad}{a^\dagger}

\newcommand{\hc}{\textrm{h.c.}}
\newcommand{\rd}{\textrm{d}}

\newcommand{\rana}{\rangle}
\newcommand{\lan}{\langle}

\newcommand{\na}{\langle n_A\rangle}
\newcommand{\nb}{\langle n_B\rangle}
\newcommand{\nc}{\langle n_C\rangle}
\newcommand{\nei}{\langle n_1\rangle}
\newcommand{\nz}{\langle n_2\rangle}
\newcommand{\nr}{\langle n_r\rangle} 
\newcommand{\ini}{\textrm{in}}
\newcommand{\im}{\textrm{im}}
\newcommand{\fin}{\textrm{fin}}

\newcommand{\emi}{\textrm{em}}

\newcommand{\inter}{\textrm{int}}
\newcommand{\rad}{\textrm{rad}}
\newcommand{\dec}{\textrm{dec}}

\newcommand{\sys}{\textrm{sys}}
\newcommand{\ext}{\textrm{ext}}
\newcommand{\tls}{\textrm{TLS}}
\newcommand{\N}{{\cal N}}

\newcommand{\lana}{\langle}
\newcommand{\hb}{\bar{h}}
\newcommand{\gb}{\bar{g}}
\newcommand{\hba}{\mathchar'26\mkern-7mu h}

\journalname{Foundations of Physics}
\begin{document}

\title{Fermi's golden rule and the second law of thermodynamics}


\author{D.~Braak         \and
        J.~Mannhart 
}


\institute{D.~Braak \and J.~Mannhart \at
              Max Planck Institute for Solid State Research\\
              70569 Stuttgart, Germany\\
              \email{d.braak@fkf.mpg.de}\\
              \email{j.mannhart@fkf.mpg.de}
}

\date{Received: date / Accepted: date}

\maketitle

\begin{abstract}
 We present a Gedankenexperiment that leads to a violation
 of detailed balance if quantum mechanical transition probabilities are
 treated in the usual way by applying Fermi's ``golden rule''. This Gedankenexperiment introduces a collection of two-level systems that
 absorb and emit radiation randomly through non-reciprocal coupling to a
 waveguide, as realized in specific chiral quantum optical systems. 
The non-reciprocal coupling is modeled by a hermitean Hamiltonian and is
compatible with the time-reversal invariance of unitary quantum dynamics.
Surprisingly, 
 the combination of non-re\-ci\-pro\-ci\-ty with probabilistic radiation
 processes entails negative entropy production. Although the considered
 system appears to fulfill all conditions for Markovian stochastic dynamics, such
 a dynamics 
 violates the Clausius inequality, a formulation of the second law of thermodynamics.  
 Several implications concerning the interpretation of the quantum mechanical formalism are discussed.
\end{abstract}

\section{Introduction}
\label{intro}
The probabilistic nature of quantum physics is related to a process whose correct interpretation and mathematically sound formulation is still under debate, the so-called collapse of the wave function \cite{miller}.
The quantum mechanical collapse
is of fundamental importance for
our common-sense concept of macroscopic reality. In most cases, collapses are invoked in the context of
measurements of  quantum observables.
But extending the applicability range of quantum mechanics beyond the
microscopic realm prompts the question whether the discontinuous change of
the wave function during the measurement is a physical process or not. 
Schr\"odinger's famous Gedankenexperiment involving a
cat demonstrates drastically the consequences of the assumption that the
collapse process is  merely epistemic and thus concerns only the
knowledge of the observer \cite{schrod}. 
Schr\"odinger's goal was to demonstrate that a paradox arises if the
collapse is not considered as a microscopic process taking place independently
from the presence of an observer.
To explain the absence of macroscopic superpositions without the assumption of
a physical collapse,
the ``decoherence'' interpretation
considers not individual systems but the
density matrix of an ensemble which becomes mixed after tracing over unobserved environmental
degrees of freedom \cite{zeh,zurek,joos}. Although this avoids the
need for a ``measurement apparatus'', the tracing operation is not physical
but epistemic. Without a physical mechanism triggering a real collapse event
as assumed \textit{e.g.} in \cite{GRW}, the
decoherence interpretation is equivalent to the many-world hypothesis
\cite{everett}.      

Already in Schr\"odinger's cat example, the transition from 
unitary and deterministic to probabilistic evolution is tied to a microscopic, unpredictable
\textit{event}, the decay of a radioactive nucleus. There are strong arguments
from a fundamental perspective supporting the occurence of spatio-temporally
localized events as objective and observer-independent
equivalent of collapse processes \cite{haag}. 

These random events, called ``quantum jumps'' in the early debate between
Schr\"o\-din\-ger and Bohr \cite{schr-qj}, are generally thought to underlie
the statistical character of the emission and absorption of light quanta by
atoms. Einstein used arguments from the theory of classical gases to
derive Planck's formula by assuming detailed balance between the atoms and the
radiation in thermal equilibrium. His derivation did not require a microscopic
Hamiltonian \cite{einstein1916}. Nevertheless, it is easy to derive the
corresponding rate equations for the time-dependent occupancy $\langle
n\rangle(t)$ of a light mode with frequency $\Om$ coupled resonantly to $M$
two-level systems (TLS) within quantum mechanics. The transition probabilities follow from the microscopic interaction Hamiltonian by employing Fermi's {\it golden rule} \cite{goldenrule,fermi}, which tacitly incorporates the collapse event by replacing the unitary time evolution by a stochastic process. 

The interaction Hamiltonian is well known \cite{c-t,loud}:
\beq
H_{\inter}=g\sum_{l=1}^M\left(a\s^+_l+\ad\s^-_l\right),
\label{hint-bb}
\eeq
where $a,\ad$ denote the annihilation/creation operators of the radiation mode
and  $n=\ad a$. The Pauli lowering/raising operators $\s^-_l,\s^+_l$ describe
the $l$-th two-level system with $H_{\tls}=\hba\Om(\sum_l\s_l^+\s_l^--1/2)$. The
probabilities for a transition between the upper and lower state of a TLS
accompanied by the emission or absorption of a photon with frequency $\Om$ are
computed with the golden rule (see section~\ref{blackbody}) to obtain the rate equations 
\begin{align}
  \frac{\rd \langle n\rangle}{\rd t} &= \g\left[\langle m\rangle(\langle n\rangle+1)- (M-\langle m\rangle)\langle n\rangle\right],
  \label{bb-n}\\
  \frac{\rd \langle m\rangle}{\rd t} &= \g'\left[(M-\langle m\rangle)\langle n\rangle -\langle m\rangle(\langle n\rangle +1)\right].
  \label{bb-m}
\end{align}
Here, $\langle m(t)\rangle$ denotes the time-dependent average number of excited TLS. The rate constants $\g,\g'$ depend on the coupling $g$ and the density of states of the radiation continuum around $\Om$ (see below).
These equations describe the irreversible change of average quantities and thus use the ensemble picture of statistical mechanics \cite{wal}. Nevertheless they account for the temporal behavior of a single system as well, because a {\it typical} trajectory will exhibit a fraction $m(t)/M$ of excited TLS close to $\langle m(t)\rangle/M$ for sufficiently large $M$ \cite{goldstein,baldovin,cerino}.

It is crucial that the rate equations \eqref{bb-n} and \eqref{bb-m} satisfy the \textit{detailed balance condition}, which implies that they lead from arbitrary initial values $\langle n\rangle(0), \langle m(0)\rangle$ to a unique steady state characterized by 
\beq
\langle m\rangle(\langle n\rangle +1) = (M-\langle m\rangle)\langle n\rangle.
\label{det-bal}
\eeq
We have the relations
\begin{align}
  P^e=\langle m\rangle/M,\quad & P^g=(M-\langle m\rangle)/M,\nn\\
  P_{e\ra g}=\g'(\langle n\rangle +1),\quad & P_{g\ra e}=\g'\langle n\rangle,
\end{align}
for the probabilities $P^g$ ($P^e$) for a TLS to be in its ground (excited) state and the probabilities for emission $P_{e\ra g}$ and absorption $P_{g\ra e}$. 
Equation \eqref{det-bal} can therefore be written as
\beq
P^eP_{e\ra g} = P^gP_{g\ra e}
\label{det-bal2}
\eeq
which is the definition of detailed balance \cite{wal,reichl}.
Equation \eqref{det-bal} 
entails Planck's formula for thermal equilibrium between radiation and matter. If one considers \eqref{bb-m} as an equation of motion for the probability $P^e(t)$, even a microscopic system consisting of a single TLS will thermally equilibrate with the surrounding continuum of radiation.

This surprising result rests on the fact that the TLS does not interact
with the light mode exactly on resonance only but with all modes in a frequency
interval of width $\D$ around $\Om$ with a similar strength $g$
\cite{c-t,loud}. The energy uncertainty $\hba\D$ allows for a radiation event
occuring during a short time span $\tau_c\sim \D^{-1}$, whereas the process
itself is energy conserving (see \cite{c-t}, p. 419). The coupling to a
continuum of modes
leads therefore to real and irreversible {\it microscopic} processes, the emission or absorption of 
light quanta, although no macroscopic measurement apparatus is involved. Such a
microscopic collapse process is tacitly assumed whenever the golden rule is
employed. If, however, the TLS is embedded into a cavity and coupled only to a single
radiation mode, the collapse cannot take place; the system shows the
Rabi oscillations of an unitarily evolving state instead, the TLS being entangled with
the bosonic mode. To this case, the golden rule cannot be applied.     
Therefore, the golden rule cannot be taken as an approximation to the full
unitary time development given by solving the Schr\"odinger equation, although
it corresponds formally to a perturbative computation of the unitary dynamics
for short times \cite{zhang}. The very concept of a \textit{transition rate} implies that
the deterministic evolution of the state vector is replaced with a
probabilistic description of events. The central element in the computation is the overlap $\langle\psi_{\textrm{final}}|H_\inter|\psi_{\textrm{initial}}\rangle$, which, according to the Born rule, determines the probability for the transition from $|\psi_{\textrm{initial}}\rangle$ to $|\psi_{\textrm{final}}\rangle$. The Born rule is invoked here although the process is not a macroscopic measurement. The ``macroscopic'' element is provided by the continuum of radiation modes. The presence of this continuum causes the necessity for a statistical description \cite{c-t}.

Within the decoherence interpretation, one would require
that the continuum of radiation modes acts as an environment for the TLS, the
``system''. The environment is then traced out to yield the irreversible
dynamics of the system alone. But if the system consists of the walls of a
hohlraum, it excercises a strong influence on the ``environment'', the
enclosed radiation, such that they equilibrate together. Therefore, it is not
possible to separate the environment from the system to explain the
microscopic collapse processes driving the compound system towards thermal equilibrium.

The rate equations neglect completely the coherences of the TLS. The
irrelevance of the coherences follows
naturally from the interpretation of the radiation process as a collapse: 
each such event projects the
TLS into their energy eigenbasis, just as a macroscopic
measurement projects any
quantum system into the basis entangled with the eigenbasis of
the measurement device \cite{neu}.
A macroscopic measurement device is not needed here because both
subsystems, the TLS and the radiative continuum, contain a macroscopic number
of degrees of freedom. This alone seems to justify a statistical description as
in classical gas theory, although the interaction Hamiltonian \eqref{hint-bb}
has no classical limit. Indeed, in our case the Born rule
replaces the assumption of  ``molecular chaos'', needed in Boltzmann's derivation of the H-theorem
\cite{wal,reichl}. Therefore, it seems almost natural that the rate equations
(\ref{bb-n}),(\ref{bb-m}) lead to thermal equilibrium from a
non-equilibrium initial state although the radiation and the collection of
atoms are both treated as ideal gases. If $\langle m(0)\rangle$ and $\langle n\rangle(0)$
correspond to equilibrium ensembles with different temperatures at $t=0$,
\beq
T_{\tls}(0)=\frac{\hba\Om}{k_B\ln\left(\frac{M}{\langle m(0)\rangle}-1\right)}, \quad
T_{\textrm{rad}}(0)=
\frac{\hba\Om}{k_B\ln\left(\frac{1}{\langle n\rangle(0)}+1\right)},
\eeq
the rate equations derived from the quantum mechanical interaction Hamiltonian \eqref{hint-bb} together with the golden rule entail thermal equilibration according to Clausius' formulation of the second law of thermodynamics: the two gases exchange heat which flows from the hotter to the colder subsystem until a uniform temperature and maximum entropy of the compound system is reached \cite{wal,reichl}.
 
Certainly, the rate equations do not correspond to the exact quantum dynamics
of the system, which is non-integrable in the quantum sense if the interaction \eqref{hint-bb} is
generalized to a continuum of bosonic modes, rendering it equivalent to the
spin-boson model \cite{spin-boson}. To obtain the exact evolution equation for
the density matrix of both the TLS and the radiative modes, one would have to
solve the full many-body problem. The golden rule is then seen as a method to
approximate the time-dependent expectation values $\langle n\rangle(t)$,
$\langle m(t)\rangle$, which is justified by a large body of experimental
evidence, but not through an analytical proof of equivalence between both
approaches. It may even be that the golden rule provides a phenomenological
description of microscopic collapse processes whose actual dynamics is not yet
known. In this case, the full quantum mechanical calculation would not yield
(\ref{bb-n}),(\ref{bb-m}), although they account correctly for the
experimentally observed dynamics. The
rate equations are derived from the interaction Hamiltonian \eqref{hint-bb} and the golden rule
(see sections~\ref{blackbody} and \ref{open}), thereby demonstrating its applicability 
in the situation considered. This corroboration of the golden rule
based on experimental results is independent of any theoretical justification
of this rule.

 The interaction Hamiltonian \eqref{hint-bb} satisfies the detailed balance
 condition which is crucial for the physically expected behavior. 
 We shall demonstrate in the next section that this condition does not follow
 from elementary principles like time reversal invariance or hermiticity as
 \eqref{hint-bb} seems to suggest. On the contrary, there are feasible
 experimental setups violating the detailed balance condition while satisfying
 all other prerequisites for the application of the golden rule. The
 consequence of this violation is a macroscopic disagreement with the second law of thermodynamics.    

\section{The  Gedankenexperiment}
\label{sec:1}
\begin{figure}
\centering
	\includegraphics[width=0.7\textwidth]{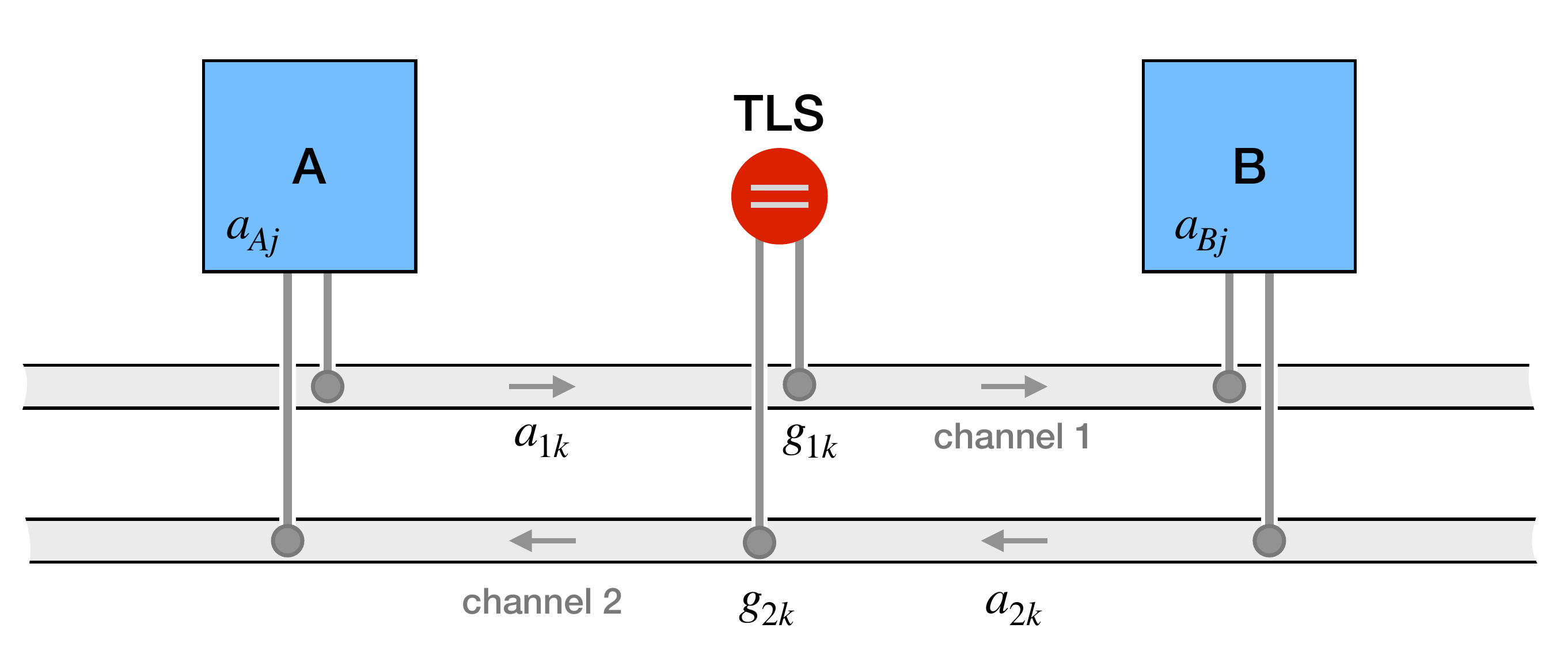}
	\caption{Layout of the Gedankenexperiment. Two reservoirs $A$ and $B$
          containing black-body radiation are coupled via a non-reciprocal,
          open waveguide to a collection of two-level systems (TLS) and to
          each other. The right-, respectively left-moving modes in channels 1
          and 2 couple with different parameters $g_1$ and $g_2$ to the two-level systems.}
        \label{Fig1}
\end{figure}

In the Gedankenexperiment, we consider two identical cavities $A$ and $B$
supporting a quasi-continuous mode spectrum described by bosonic annihilation
operators $a_{Aj}$, $a_{Bj}$ and frequencies $\om_j$ (Fig.~\ref{Fig1}). They are
coupled bilinearly to the right- and left-moving modes $a_{1k}$, $a_{2k}$ of
an open-ended waveguide  which form  a quasi-continuum like the cavity modes
\cite{dayan}.
 The loss processes through the open ends of the waveguide are
 caused formally by a heat bath at $T=0$, leading to a reduction of the system
 entropy through heat transport. This coupling to the outside world is one
 argument for the applicability of the golden rule, the second is the already discussed
 quasi-continuum of modes. But even if the full continuum of radiation modes in the cavities is treated as part of
 the ``system'' which would then be subject to purely unitary
 time evolution, the waveguide would still couple to a decohering
 ``environment'', justifying the statistical description even if one denies that
 real collapse events take place in the system itself. Here we study a model which is commonly used in quantum optics to describe unidirectional loss processes \cite{haroche}. In this way, the generally accepted arguments substantiating irreversible evolution equations can be transferred to the present situation.
 
The modes $a_{1k}$ and $a_{2k}$ of the waveguide are coupled to a collection of $M$ two-level systems located at the center of the waveguide (see Fig.~\ref{Fig1}).
The total Hamiltonian is given by
\beq
H=H_A + H_B + H_{wg} + H_{\tls} + H^1_{\inter} +  H^2_{\inter}.
\label{ham}
\eeq
$H_q$ denotes the Hamiltonian in cavity $q$ for $q=A,B$,
\beq
H_q=\hba\sum_{j}\om_j a_{qj}^\dagger a_{qj}.
\eeq
The Hamiltonian of the waveguide reads
\beq
H_{wg}=\hba\sum_{k}\om_k \left(a_{1k}^\dagger a_{1k} + a_{2k}^\dagger a_{2k}\right), 
\eeq
where the modes 1 and 2 belong to waves traveling to the right and to the left, respectively. For the TLS we have $H_{\tls}=(\hba\Omega/2)\sum_{l=1}^M\s^z_l$, with the Pauli matrix $\s^z$. The coupling between the reservoirs and the modes 1 and 2 of the waveguide is bilinear,
\beq
H^1_{\inter}=\sum_{q=A,B}\sum_{j,k} h_{jk}\left(a_{qj}^\dagger [a_{1k} + a_{2k}] +\hc\right).
\label{hint1}
\eeq
Using the rotating wave approximation, the interaction with the TLS has the
standard form \cite{c-t,loud} which is equivalent to \eqref{hint-bb},
\beq
H^2_{\inter}=\sum_{l=1}^M \left(\sum_{k} g_{1k}a_{1k} +g_{2k}a_{2k}\right)\s^+_l + \hc,
\label{hint2}
\eeq
where $\s^+_l$ denotes the raising operator of the $l$th TLS.
We consider in the following the (time-dependent) average occupancy per mode
$j$, $\lana n_q\rana(t)$ for reservoir $q=A,B$ in an energy interval around
the TLS energy, $\Om-\Delta/2 <\om_j<\Om+\Delta/2$, where $\D$ is much larger than the natural linewidth of spontaneous emission from an excited TLS into the waveguide. The occupancy does not depend on $j$ if the couplings $h_{jk}$, $g_{(1,2)k}$,  the density of states $\rho_q(\hba\om_j)$ of the reservoirs and $\rho_{1,2}(\hba\om_k)$ of the wa\-ve\-gui\-de are constant in the frequency interval of width $\Delta$ around $\Om$.

It is crucial that  $g_{1k}\neq g_{2k}$, which specifies that the TLS couple
with unequal strengths to the right- and left-moving photons in the waveguide.
Such an unequal coupling is
a hallmark of chiral quantum optics \cite{lodahl}. Although the interaction
Hamiltonian \eqref{hint2} appears to break time-reversal invariance, as the
time-reversal operator maps left-moving to right-moving modes, this is
actually not the case because the effective interaction term \eqref{hint2} does
not contain the polarization degree of freedom. The angular momentum selection
rules for light-matter interaction lead naturally to a dependence of the
coupling strength on the propagation direction in engineered geometries
\cite{luxmoore}, especially if the spin-momentum locking of propagating
modes in nanofibers is employed \cite{junge,shomroni}. The unwanted coupling
to non-guided modes can be effectively eliminated, leading to large
coupling differences
$|g_{1k}-g_{2k}|$ \cite{soellner,lefeber}.  

We shall now study the temporal behavior of the two cavities, assuming at time
$t=0$  separate thermal equilibria in $A$, $B$ and the TLS system, all at
the same temperature $T$. The probability $\langle m(0)\rangle/M$ for a TLS to be
excited obeys the Boltzmann distribution
\beq
\frac{\langle m(0)\rangle}{M}=\frac{1}{e^{\hba\Om/k_BT}+1}.
\label{boltz}
\eeq
Considering the $M$ TLS as independent classical objects, their Gibbs entropy 
is given by
\beq
S_M=-k_BM\big(p_e\ln p_e +(1-p_e)\ln (1-p_e)\big),
\label{entropyTLS}
\eeq
with $p_e(t)=\langle m(t)\rangle/M$. This approach is justified by the quick relaxation of
the two-level systems by non-radiative processes, which decohere them on time
scales much shorter than the time scale of spontaneous emission and quickly quench finite coherences of the TLS \cite{brouri}. This
argument for a classical description of the TLS is independent from the
general justification of the golden rule via the mode continuum
discussed above. 
Equation
\eqref{entropyTLS} provides an upper bound for the entropy of the TLS subsystem \cite{lanford}. Analogously, the entropy of the compound system reads
\beq
S^{\sys}=S_A +S_B + S_M,
\label{sys-entropy}
\eeq
where $S_q$ denotes the v.~Neumann entropy of the radiation in cavity $q$ for $q=A,B$.
The average occupation number $\lana n_q(\om)\rana$  per mode at frequency $\om$ follows from the Bose distribution
\beq
\lana n_q(\om)\rana(0)=\frac{1}{e^{\hba\om/k_BT}-1}.
\label{n-T}
\eeq
\par
Because the temperature depends on $\lana n_q\rana$ as described by \eqref{n-T},
we can define effective temperatures $T_q(t)$ by $\lana n_q\rana(t)$ for each reservoir $q$ under 
the assumption that the photons in each reservoir thermalize in the usual way quickly as a non-interacting Bose gas. 
Furthermore, we consider the case that $\lana n_{(1,2)k}\rana(t)=0$ for the occupancy of the modes in the waveguide, i.e., the waveguide is populated through the sufficiently weak coupling to the reservoirs and its modes appear only as intermediate states (see section~\ref{open}).
Neglecting the waveguide, the system is thus composed of three subsystems,
each in thermal equilibrium at any time $t>0$, with locally assigned
time-dependent temperatures. The subsystems interact through random emission
and absorption processes which do not lead to entanglement, because in  each
such process the wave function undergoes a collapse towards a product
state. This description is in obvious accord with the derivation of Planck's
law given by Einstein \cite{einstein1916}. Note that the compound system is
not coupled to several thermal baths which have different temperatures. In
such a case, a description with separate
master equations for the subsystems is inconsistent if the interaction between subsystems is
still treated quantum mechanically. Using such a description, a violation of the
second law has been deduced \cite{capek1,levy}, which is only
apparent and caused by the inconsistent computation \cite{levy}.
Our system differs from those models because
the system dynamics is not described by a unitary evolution as in
\cite{capek1,levy}, but by a random process, with all subsystems coupled to
the same bath (the open waveguide).

\begin{figure}
	\centering
	\includegraphics[width=0.7\textwidth]{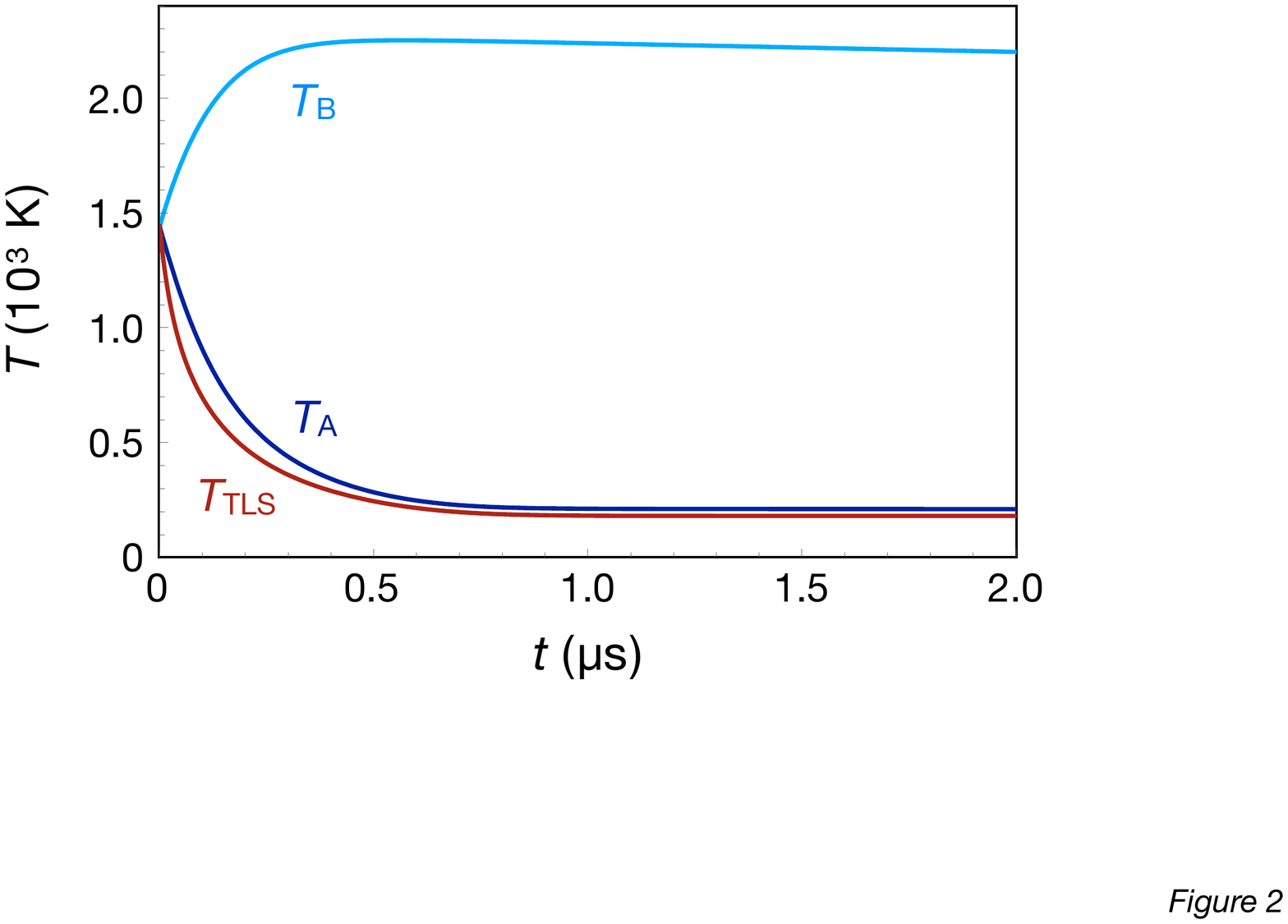}
	\caption{Solutions of the rate equations \eqref{rateA}-\eqref{rateM} as function of time, starting from initial thermal equilibrium. The effective temperatures of the reservoirs $A$ and $B$ deviate. The temperature drop of the TLS parallels that of $A$. Parameters used are
     $\g_\dec=\g_0=10$\;kHz, $\gt_{11}=\gt_1=10$\;MHz, $\g_1=100$\;kHz; $\gt_{12}=\gt_2=\g_2=0$ and $\hba\Om/kT(0)=1$. $\Om$ corresponds to a wavelength of 10\;$\mu$m.}
        \label{temp-short}
\end{figure}

\begin{figure}
	\centering
	\includegraphics[width=0.7\textwidth]{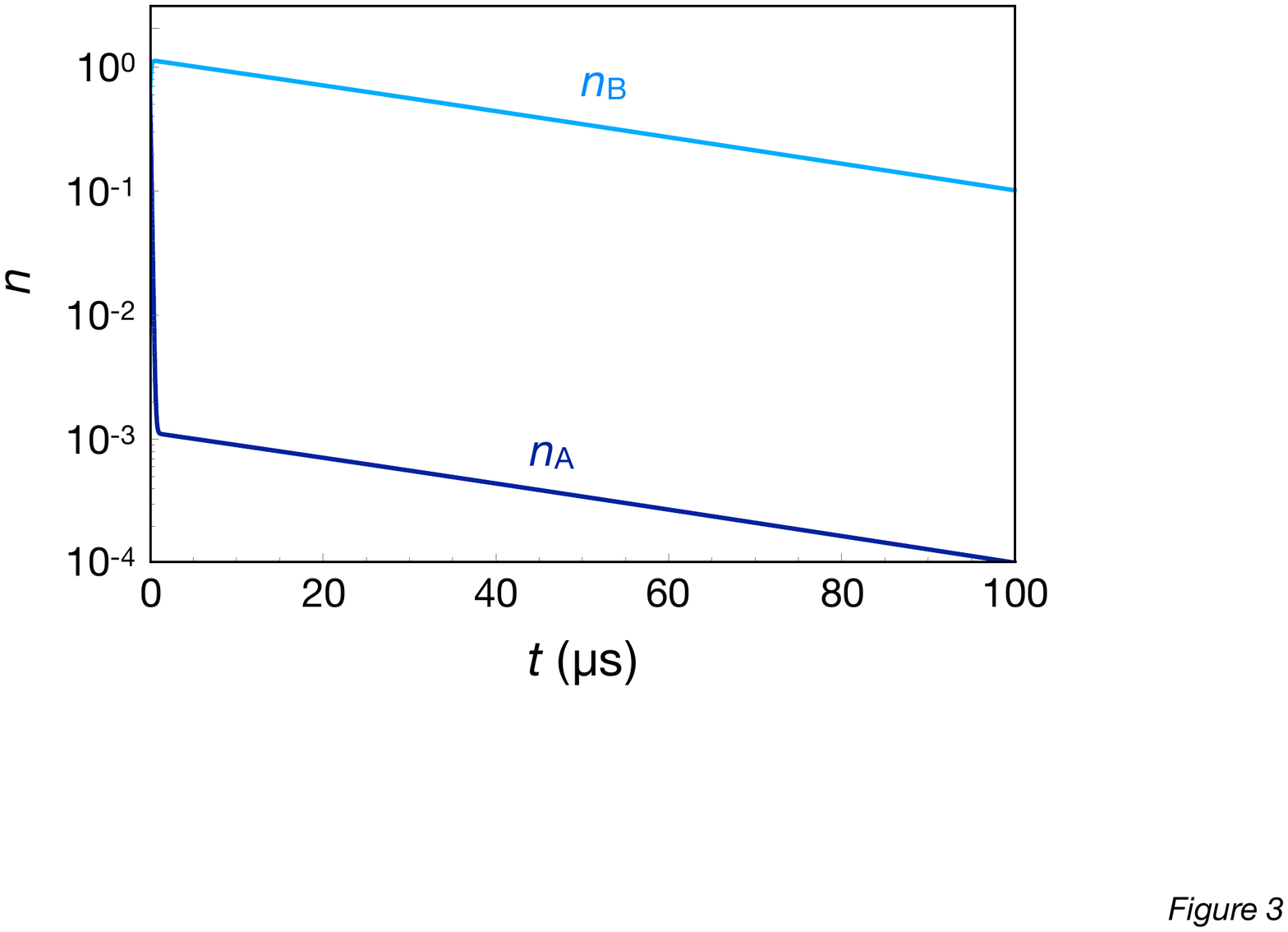}
	\caption{The occupations of reservoirs $A$ and $B$ for long times, plotted on a logarithmic scale. The ratio $\na/\nb$ is asymptotically time-independent. The losses through the open waveguide are much slower than the breakdown of local equilibrium caused by the non-reciprocal interaction with the TLS.}
	\label{occ-long}
\end{figure}

The golden rule applied to the Hamiltonian \eqref{ham} yields 
the rate equations for  $\lana n_q\rana(\Om,t)$ and $\langle m(t)\rangle$,
  \begin{align}
    \frac{\rd \na}{\rd t} &= -2\g_\dec\na + \g_0(-\na+\nb)- \g_1(M-\langle m\rangle)\na\nn\\
    &+ \g_2\langle m\rangle(\na+1),
\label{rateA}\\
\frac{\rd \nb}{\rd t} &= -2\g_\dec\nb +  \g_0(-\nb+\na)- \g_2(M-\langle m\rangle)\nb\nn\\
&+ \g_1\langle m\rangle(\nb+1) ,
\label{rateB}\\
   \frac{\rd \langle m\rangle}{\rd t}&= -[\gt_{11}(\Om) +\gt_{12}(\Om)]\langle m\rangle + \gt_1(\Om)\left[\na(M-\langle m\rangle)-(\nb+1)\langle m\rangle\right] \nn\\ 
&+\gt_2(\Om)\left[\nb(M-\langle m\rangle)-(\na+1)\langle m\rangle\right].
\label{rateM}
\end{align}
The first terms on the right hand side of \eqref{rateA} -- \eqref{rateM}
describe the loss of photons through the open ends of the waveguide. These terms are of first order in $|h_{jk}|^2$, resp. $|g_{1k}|^2$, $|g_{2k}|^2$. The following terms correspond to coherent processes of second order in the couplings. 
The effective rates $\gamma_{\dec,0,1,2}$ and $\gt_{1r},\gt_{1,2}$ used for the numerical solution of \eqref{rateA} -- \eqref{rateM} shown in Figs.~\ref{temp-short} and \ref{occ-long} belong to the strong coupling regime of the TLS and the wa\-ve\-gui\-de with values accessible within a cavity QED framework \cite{rosen}.
The chiral nature of the coupling, $g_{1k}\neq g_{2k}$, entails
$\g_1\neq\g_2$. In our example we have assumed $\gt_2=\g_2=\gt_{12}=0$, i.e.,
channel 2 is not coupled to the TLS. One sees from \eqref{rateA} --
\eqref{rateM} that the chiral coupling leads to a breakdown of the detailed
balance condition in second-order processes because absorption is no longer
balanced by stimulated and spontaneous emission. The radiation processes generate an effective transfer
of photons from reservoir $A$ to $B$ on a time scale $\tau_{\textrm{char}}$ given by the strong
coupling between channel 1 and the TLS, $\tau_{\textrm{char}}\sim\gt^{-1}=0.1\;\mu$s. This corresponds to a difference in the local temperatures calculated via \eqref{n-T}.
Figure~\ref{temp-short} shows the temporal behavior of the temperatures of reservoirs $A$, $B$ and the TLS for intermediate times.
Although both reservoirs $A$ and $B$ loose photons through the open waveguide,
the ratio between $\na$ and $\nb$ attains a constant value for $t\rightarrow
\infty$, which is depicted in Fig.~\ref{occ-long}. This figure reveals that the losses of reservoir $B$ set in at a much later time than the spontaneous population of $B$ through reservoir $A$ and the TLS.
\section{Conflict with the second law of thermodynamics}
\label{conflict}
Due to the steady loss of photons, our system is always out of equilibrium and the only steady state solution of the equations \eqref{rateA} -- \eqref{rateM} corresponds to empty cavities and all TLS in their ground state. It is clear that the system entropy\footnote{For our purposes it plays no role whether the Boltzmann or the Gibbs/v.~Neumann entropy is employed, because all definitions coincide in equilibrium and we assume {\it local} thermal equilibrium for all three subsystems, the two cavities and the TLS, albeit with possibly differing local temperatures.} diminishes at any time due to the outgoing heat flow. The question arises how to apply the second law of thermodynamics to this situation.
The second law has been formulated in several versions (see, {\it e.g.}, \cite{planck1,lieb,uffink}). It is not the purpose of this paper
to discuss these in detail.
Three representative formulations provide exemplary definitions of the second law \cite{wal,reichl}:\\
\begin{tabular}{lp{0.8\textwidth}}
(1)&  The entropy of the universe always increases.\\
(2)&  The entropy of a completely isolated system stays either constant or increases.\\
(3)&  The entropy of a system thermally coupled to the environment satisfies Clausius' inequality: $\D S \ge \D Q/T$.
\end{tabular}\par
\noindent
Variant (1) is not subject to our Gedankenexperiment, because the entropy production including the environment is formally infinite, as the external bath has zero temperature. Variant (2) follows from variant (3) because the heat transfer $\D Q$ vanishes. Variant (3) applies to the present system: the heat transfer $\D Q$ is negative and therefore also $\D S$ may be negative. However, the second law in the form of Clausius' inequality forbids the case $\D S \le \D Q/T$: The local entropy production $\s$ in the system
must be non-negative \cite{reichl}.

\begin{figure}
	\centering
	\includegraphics[width=0.7\textwidth]{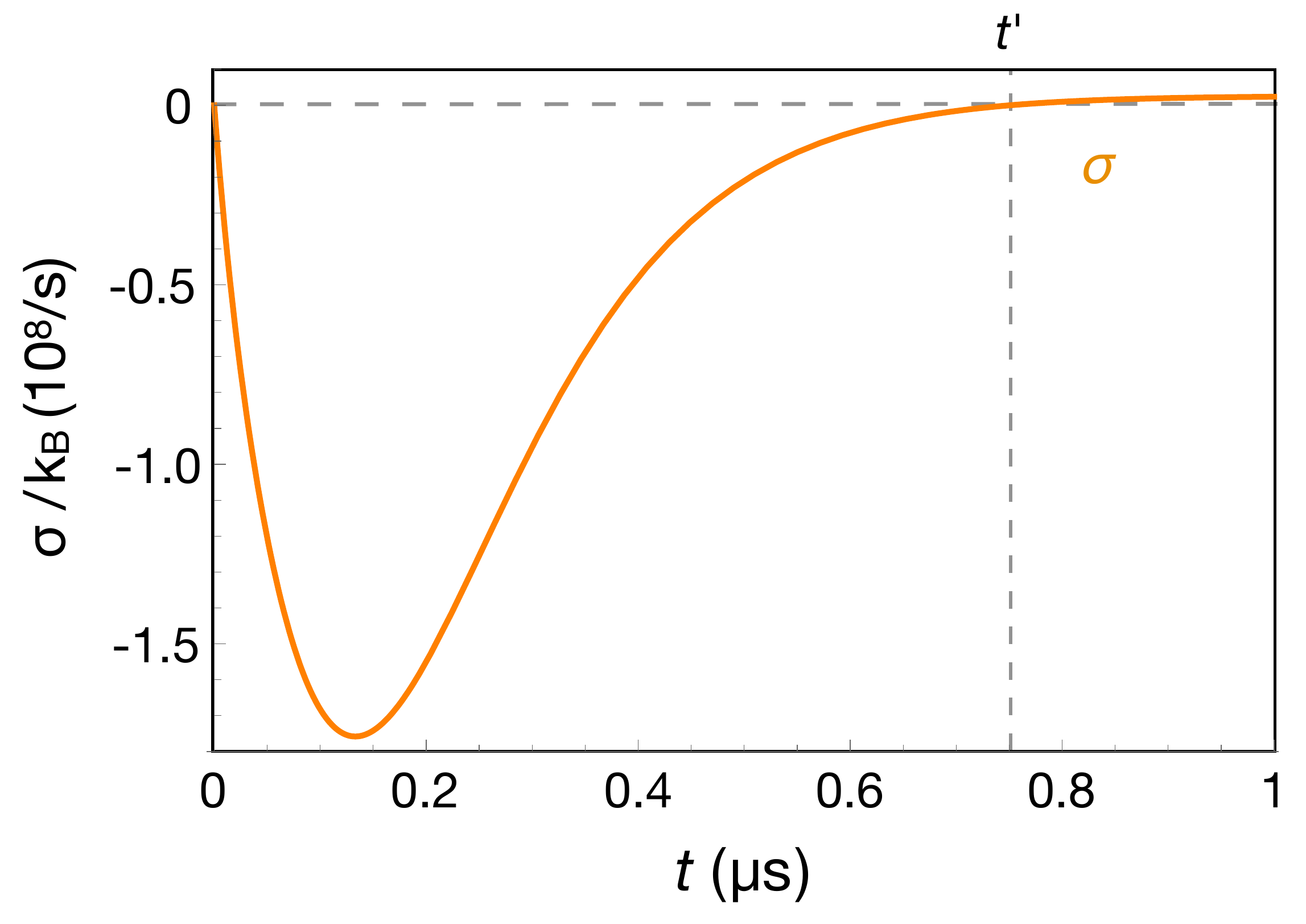}
	\caption{The total entropy production $\s(t)$ within the system calculated for $\N\g_0=1$\;MHz,
          $\gt_1=10$\;MHz, $\gt_2=0$. For these parameters, the entropy production is negative until $t'\sim 0.75\;\mu$s, when it turns positive and stays so. This behavior entails that for $t<t'$ the second law of thermodynamics is not satisfied. Note that $t'$ is appreciably larger than $\tau_{\textrm{char}}$.}
	\label{sigma}
\end{figure}

The change of system entropy $S^{\sys}(t)$ can be written as \cite{spohn}
\beq
\frac{\rd S^{\sys}}{\rd t} = -\int\rd{\bm o}\cdot{\bm J}_{\sys} +\s,
\label{entropyrate}
\eeq
where the surface integral of the entropy current ${\bm J}_{\sys}$ accounts for the heat transfer to the environment, characterized by the rates $\g_\dec$ and $\gt_{11},\gt_{12}$.  We find for $\s(t)$ (see section \ref{entr-prod}) 
\beq
\s(t)= \s_{A,B}(t) + \s_{A,\tls}(t) + \s_{B,\tls}(t),
\label{entropyprod}
\eeq
with
\begin{align}
  \s_{A,B} &=k_B{\cal N}\g_0(\nb - \na)\big(\ln(\nb[\na+1])-\ln(\na[\nb+1])\big),
  \label{sab}\\
  \s_{A,\tls}&=k_B\big(\gt_2\langle m\rangle(\na+1)-\gt_1\na(M-\langle m\rangle)\big)\big(\ln(\langle m\rangle[\na+1]) \nn\\
  &-\ln([M-\langle m\rangle]\na)\big),\label{sat}\\
  \s_{B,\tls}&=k_B\big(\gt_1\langle m\rangle(\nb+1)-\gt_2\nb(M-\langle m\rangle)\big)\big(\ln(\langle m\rangle[\nb+1]) \nn\\
  &-\ln([M-\langle m\rangle]\nb)\big),\label{sbt}
\end{align}
with ${\cal N}=\gt_1/\g_1=\gt_2/\g_2$.
The three contributions result from the heat exchange between the three subsystems. Only $\s_{A,B}$ is always non-negative, because it has the form $(x-y)(\ln(x)-\ln(y))$ characteristic for systems satisfying the detailed balance condition. Because $\gt_1\neq\gt_2$, the two other contributions are not necessarily non-negative. Fig.~\ref{sigma} shows $\s(t)$ calculated using the parameters of Fig.~\ref{temp-short}.
The entropy production is negative for $t<0.75\;\mu$s. The time span during which the entropy is reduced beyond the loss to the environment is almost an order of magnitude longer than the characteristic time scale
$\tau_{\textrm{char}}= 0.1\;\mu$s. The especially notable point is the entropy reduction which occurs although 
the dynamical evolution
started with thermal equilibrium between the subsystems.

During the exchange of heat with the other subsystems,
each subsystem remains in local thermal
equilibrium. In principle, the exchange of heat between subsystems gives only a lower bound to the entropy production for irreversible processes and the actual $\s(t)$ could be larger than the value given in \eqref{entropyprod}, which contains the contributions from the mutual heat exchange. However, the entropy change in each subsystem can be computed directly via the formulae \eqref{entropyTLS} and \eqref{entropy2}, \eqref{entropy3} as well. Doing so, one finds that no additional entropy production besides the mutual heat exchange is generated in the process described by the rate equations \eqref{rateA} -- \eqref{rateM}. This, obviously, is due to the fact that during this process each subsystem remains in local thermal equilibrium. 
It follows that the entropy change of the full system would 
even be negative if the initial entropy of the TLS subsystem was lower
than the upper bound given in \eqref{entropyTLS}, because the entropy change is caused solely by mutual heat transfer between subsystems if the condition of local equilibrium is satisfied.

We conclude that the initial thermal equilibrium between reservoirs and the TLS 
is unstable and variant (3) of the second law is violated. This is true although the total
entropy of the system plus the environment always increases. The second law,
applied to the system alone, demands a non-negative local entropy production
for any process driven by the coupling to the bath at $T=0$
\cite{reichl,spohn}. Such processes may generate local temperature gradients
between the reservoirs, but $\s(t)$ must always be larger or equal zero to satisfy Clausius' inequality. The violation of this inequality in our Gedankenexperiment shows clearly that the chiral coupling between TLS and the waveguide generates radiation processes which are in conflict with thermodynamics if they are treated statistically in the same way as black body radiation.     

\section{Discussion and Conclusions}
\label{sec:2}
In discussing the experiment we first note that it is based on phenomena taking place in the realm
where the quantum world interfaces classical physics.
The three subsystems are clearly macroscopic, but their interaction Hamiltonian (\ref{hint1},\ref{hint2}) is purely
quantum mechanical and cannot be treated in a (semi-)classical approximation.
 A unique quantum feature of the interaction is given by the fact that the emission rate of the TLS depends on the occupation of the \textit{final} states. In section~\ref{embed}, we demonstrate that this counter-intuitive effect leads to the restoration of detailed balance in a cavity system without non-reciprocal elements.
The non-unitary, probabilistic state development of the device can neither be achieved in classical Hamiltonian dynamics nor in the unitary pure quantum regime described by the Schrödinger equation. The collapse processes that link the classical world and the quantum regime \cite{Leggett2005}
are the cause of the thermal imbalance between the otherwise equivalent reservoirs $A$ and $B$.
Our Gedankenexperiment therefore reveals a clear conflict between thermodynamics and the
probabilistic description of quantum phenomena on a macroscopic scale.
The corresponding rate
equations \eqref{rateA} -- \eqref{rateM} do not satisfy the condition for detailed balance.
Instead they predict a time-dependent state that violates 
the Clausius inequality, {\it i.e.} the original formulation of the second law of thermodynamics.

According to the quantum description of the statistical absorption and emission processes, the chirally coupled cavities are expected to
develop unequal occupation numbers. This imbalance creates  a temperature gradient
between them, although no work is done on the system, which is coupled to the
environment through the open waveguide only. This coupling to a heat bath at
temperature zero leads to heat flow out of the system which is usually
accompanied by a positive local entropy production. But in our case the
entropy production within the system is negative during a well defined time
interval. This interval is larger than the time characterizing the coupling between the subsystems. The violation of the second law is temporary. Because this violation is described by the rate equations, it is not caused by a statistical fluctuation. Such a fluctuation may occur in the stochastic evolution of the state vector of a single system, even when the initial state is typical \cite{jarz2}, but cannot appear in the fully deterministic equations for averages.

The rate equations \eqref{rateA} -- \eqref{rateM} have been derived under the
assumption that the two channels of the waveguide are fed by $A$ and $B$
through the emission of wavepackets which in turn interact with the TLS in a
causal fashion. The emission and absorption of single photons by the TLS are
considered thus as probabilistic {\it processes} taking place within a finite
time span, due to the quasi-continuum of  modes available in the waveguide and
the reservoirs. They therefore satisfy causality: It is not possible for a
right-moving photon in channel 1 to be emitted by the TLS  and be subsequently
absorbed by reservoir $A$. The photons entering $A$ are either generated by a
fluctuation in channel 1 of the open waveguide  or arrive through channel
2. In the latter case, they may come from the outside, from  a TLS or from reservoir $B$. Pure scattering events at the TLS are neglected in this approximation because they are of higher order in the coupling constants.  Their inclusion cannot restore the detailed balance broken by the chiral coupling.

Our reasoning is based on the assumption that the interaction of the TLS with
the radiation continuum leads to \textit{real events} \cite{haag} which must
be described statistically, and are therefore caused by a collapse
process. The physical mechanism of this ``real" collapse  plays no role in
these considerations because no hypothesis beyond the golden rule enters the
derivation of the rate equations. Of course, we have also assumed that the
macroscopic nature of the radiation and the collection of TLS removes any detectable
entanglement between the subsystems. It entails the statistical
independence of the radiation processes and therefore Markovian
dynamics\footnote{A Markovian master equation yields transition probabilities in
  accord with the golden rule \cite{alicki}.}.
This assumption is corroborated by all available experimental evidence up to
now. In case the second law of thermodynamics would be correct and therefore the
presented statistical analysis wrong, the validity of the second law  would be tantamount to the actual
realization of a macroscopic superposition of states in the cavity system,
although it is coupled to an unobserved environment, the open ends of the waveguide.  

\begin{figure}
	\centering
	\includegraphics[width=0.7\textwidth]{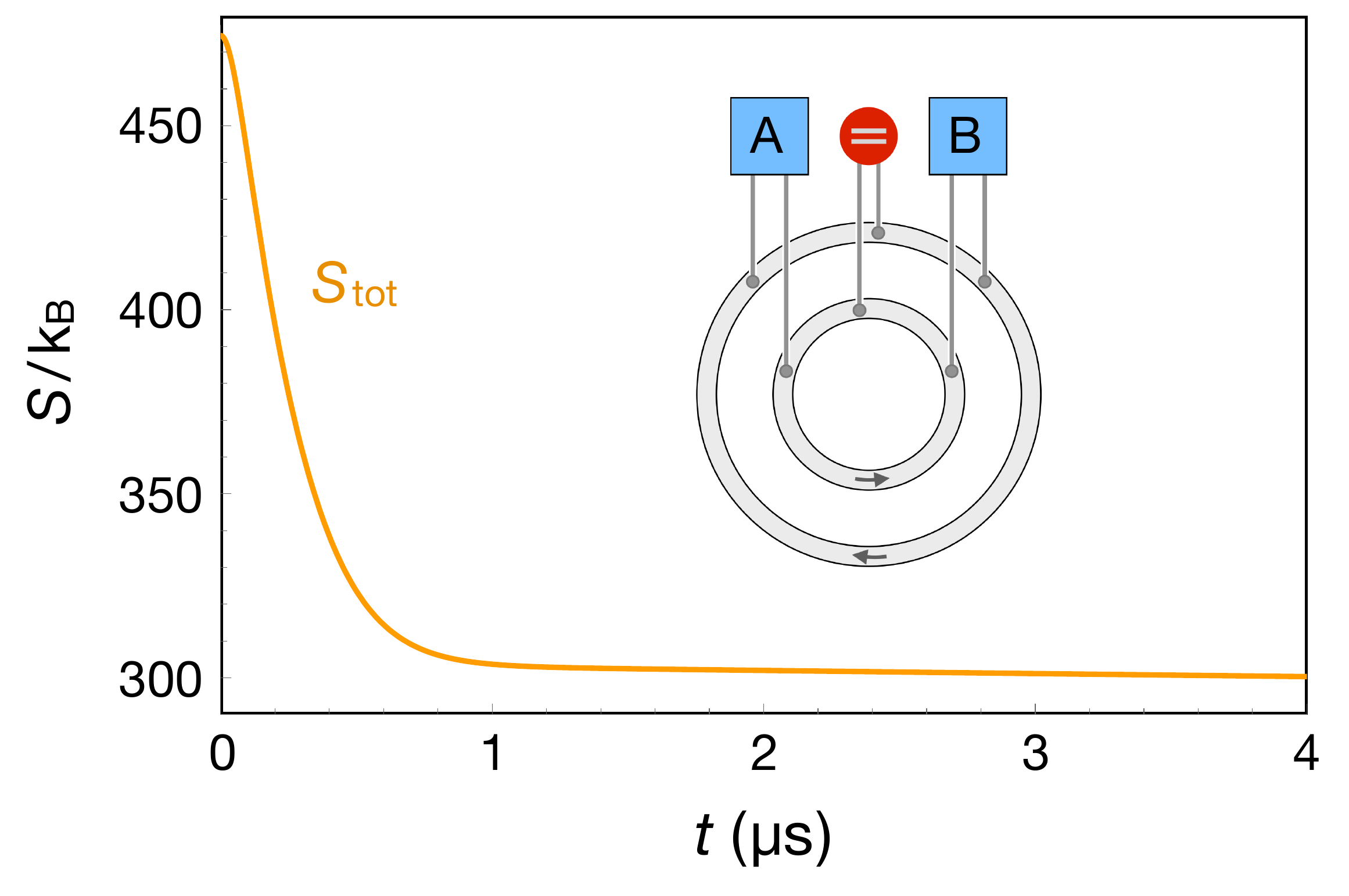}
	\caption{Temporal development of the total entropy for the closed
          variant of our system shown in the inset. The initial state at $t=0$
          (thermal equilibrium between all subsystems, including the waveguide) maximizes the entropy. The stable steady state for long times has a lower entropy.}
	\label{closedS}
      \end{figure}

Nevertheless, the use of the golden rule can also be
justified even in a completely isolated system. If the system is isolated, it could in principle be
described by the full unitary dynamics, leading to a trivial reconciliation
with the second law in its restricted form: the fine-grained Gibbs/v.~Neumann entropy does not change at all.
However, also in this case an ``environment" is present which
consists of the infinitely many degrees of freedom of the photon gas, 
decohering the dynamics of the TLS, at least
according to the opinion of the majority of physicists working in quantum optics \cite{c-t}. The effective coarse-grained description
of the closed system proceeds again via the golden rule (see
section~\ref{closed}). A similar temperature difference  between $A$ and $B$
appears and  a new steady state develops from initial thermal equilibrium for
$t\rightarrow\infty$, having a lower entropy than the initial state, thus violating variant (2) of the second law.
This is shown in Fig.~\ref{closedS}.
     The temperature difference corresponds to a ``sorting'' between the
     reservoirs $A$ and $B$ in the closed system and resembles the action of a Maxwell demon \cite{rosello}.
No information is processed, stored or erased, neither in the TLS nor
in the waveguide \cite{norton}, therefore the usual arguments for positive
entropy production based on information theory \cite{landauer,ben} cannot be
applied.

It has been argued that it is not possible to discern experimentally an
interpretation of quantum mechanics based on probabilistic dynamics and real
collapse from the decoherence interpretation which replaces the physical
collapse by an epistemic operation: the tracing over environmental degrees of
freedom in the full density matrix at the final observation
time $t_\fin$ \cite{leggett2}.
As mentioned in section~\ref{intro}, it is not known whether the photon densities in the reservoirs at $t_\fin$, calculated with the tracing procedure, would differ from the results of section~\ref{open} based on the golden rule.
If so, our proposed experiment, if performed with an isolated system, would
allow to decide between interpretations based on real collapses and epistemic interpretations. Only the latter do not contradict the second law of
thermodynamics, provided an actual solution of the full many-body problem
would effectively restore the detailed balance condition in the statistical
description.  Such a solution would also be necessary to  identify possible
reasons for the breakdown of well-established tools like the golden rule or
the Markov approximation in case the system shows the equilibrium
state predicted by thermodynamics at all times. In any case, a difference between the full solution and the approximation by the golden rule would entail another mystery: why is the approximation valid for black-body radiation in arbitrary cavities (see section~\ref{embed}) but not for chiral waveguides?

As a computation of the full quantum dynamics appears out of reach at present, the question can only be decided experimentally. A direct implementation of our model 
appears feasable with current technology \cite{dayan,lodahl,rosen}.

In case that the experiment reported unequal distributions in $A$ and $B$ for
the closed system, one would be forced to conclude that statistical processes
such as spontaneous emission and absorption are able to reduce the total
entropy  for arbitrary large isolated systems and on average, not only for short
times and small systems as expected from fluctuations \cite{jarz2}. Then the state of
classical thermal equilibrium with maximal entropy is unstable and the system
moves to steady states with a lower entropy.
In this case, entropy-reducing processes would be expected to occur actually
in nature, in structures differing greatly from our device\footnote{
Collapse processes can be provided in principle by any inelastic process, be
it scattering or (radioactive) decay. The photon reservoirs could be replaced  by any incoherent source of particles or
  energy, e.g., resistors with Johnson-Nyquist noise.
Devices closely related to the presented chiral waveguide, but being based on
coherence filters, have been presented in
\cite{Mannhart20182,mannhart_lossless_2019}. A possible solid-state realization
employing asymmetric quantum rings featuring inelastic scattering works
similarly to our quantum-optical implementation and would pump electrons
instead of photons \cite{Mannhart20181,bredol2}.}.

The contradictions presented are rooted in the still unresolved status of the
measurement problem of quantum physics: quantum mechanical probabilities can only
be computed under the assumption that a collapse (either real or epistemic)
takes place. These stochastic probabilities are formally encoded in the Born rule.
 To our
knowledge, neither the Born rule nor the golden rule have ever been used
for a derivation of the second law  of thermodynamics \cite{goldstein}. 
There have been attempts to
deduce the second law from quantum mechanics by employing
several ``coarse-graining'' prescriptions. The proof of the H-theorem given by
v.~Neumann employs assumptions about the density matrix of pure states and
macroscopic distinguishability but excludes explicitly collapse or measurement
processes from the quantum dynamics. In v.~Neumann's approach the quantum dynamics stays always unitary
\cite{neumann-H,goldstein2}. This may seem surprising,
because the collapse processes underlying the golden rule are inherently
irreversible, as noted first also by v.~Neumann \cite{neu}. To our opinion, the intrinsic probabilistic features of quantum mechanics encoded in the Born rule add an elementary irreversible process to the dynamical laws of nature. This process, the non-unitary collapse of the wavefunction, appears during macroscopic measurements but also as {\it microscopic event}, thereby leading to the correct statistics of a photon gas interacting with matter. Therefore, we disagree with the position put forward in \cite{lebowitz93}, that the irrversibility of the measurement is just due to the macroscopic nature of the apparatus and has essentially the same origin as the irreversible behavior of macroscopic variables in classical mechanics which obeys the second law of thermodynamics. We have shown that the presence of microscopic collapse processes may lead under certain circumstances to a conflict with this law.

In conclusion, transition rates of quantum systems are commonly calculated with
great success by using Fermi's golden rule. This approach is widely accepted,
as the golden rule directly results from the Born rule.
Here, we have introduced practically realizable, open and closed quantum
systems of coupled cavities and determined their behavior by applying the
golden rule. The predicted behavior of both systems violates the second law of thermodynamics.
We therefore conclude that\\
1)\ the statistical description of quantum mechanical
transitions given by the golden rule is incorrect or\\
2)\ the second law of
thermodynamics is not universally valid.

\begin{acknowledgements}
 The authors gratefully acknowledge an outstanding interaction with T. Kopp, support by L. Pavka, and also very helpful discussions with and criticism from J. Annett, E. Benkiser,  H. Boschker, A. Brataas, P. Bredol, C. Bruder, H.-P. B\"uchler, T. Chakraborty, I.L. Egusquiza, R. Fresard, 
  A. Golubov, S. Hellberg, K.-H. H\"ock, G. Ingold, V. Kresin, L. Kürten, G. Leuchs, E. Lutz, F. Marquardt, A. Morpurgo,  A. Parra-Rodriguez, H. Rogalla, A. Roulet, M. Sanz, P. Schneewei{\ss}, A. Schnyder, C. Schön, 
  E. Solano, R. Stamps, R. Valenti, J. Volz  and R. Wanke.
 D.B. thanks Raymond Fresard for the warm hospitality during his stay at CRISMAT, ENSICAEN, which was supported by the French Agence Nationale de la Recherche, Grant No. ANR-10-LABX-09-01 and the Centre National de la Recherche Scientifique  through Labex EMC3.
\end{acknowledgements}

%
\section*{Conflict of interest}
The authors declare that they have no conflict of interest.

\section{Appendix}
\label{app}
\subsection{Derivation of the rate equations \eqref{bb-n} and \eqref{bb-m}}
\label{blackbody}
The simple, albeit nonlinearly coupled rate equations \eqref{bb-n}, \eqref{bb-m} underlie Einstein's famous derivation of Planck's law for black-body radiation \cite{einstein1916}. In this section, we describe the approximations and statistical assumptions leading to the closed system of equations \eqref{bb-n}, \eqref{bb-m}.

We begin by describing the quantum state of a single system within the statistical ensemble. A collection of $M$ two-level systems with energy splitting $\hba\Om$ interacts with radiation modes $a_j$, characterized by their frequencies in the range $[\Om-\D/2,\Om+\D/2]$ and additional parameters such as momentum and polarization, summarized in the index $j$. The full many-body quantum state $|\Psi(t)\rangle$ at time $t$ is then described as
\beq
|\Psi(t)\rangle = |s_1,\ldots s_M;\{n_j\}\rangle,
\label{fullstate}
\eeq
where the $s_l=0,1$ correspond to the ground and excited state $|g_l\rangle$ and $|e_l\rangle$ of the $l$-th TLS and $n_j$ is the number of photons in mode $j$. While the states $\{\Psi\}$ form a complete basis in the full Hilbert space, we assume that at any time the system can be described by exactly one of these states, {\it i.e.} we neglect the coherence of each TLS and coherent superpositions between several TLS. We also consider the radiation field as thermal, namely each state being diagonal in the eigenbasis of the non-interacting Hamiltonian
\beq
H_0=\frac{\hba\Om}{2}\sum_{l=1}^M\s^z_l +\hba\sum_j\om_j\ad_j a_j.
\eeq
The interaction between the TLS and the radiation via the Hamiltonian \eqref{hint-bb} is incorporated as a stochastic process using the golden rule. It gives the probabilities to transition from state $|\Psi(t)\rangle$ to states
$|\Psi'(t+\del t)\rangle$ (emission) or $|\Psi''(t+\del t)\rangle$ (absorption),
\begin{align}
  |s_1,\ldots,1_l,\ldots  s_M;\{n_j\}\rangle
  &\ra  |s_1,\ldots,0_l,\ldots s_M;\{n_{j\neq k}\},n_k+1\rangle,
  \label{emiss}\\
|s_1,\ldots,0_{l'},\ldots  s_M;\{n_j\}\rangle
&\ra  |s_1,\ldots,1_{l'},\ldots s_M;\{n_{j\neq k}\},n_k-1\rangle.
\label{absorp}
\end{align}
The rate $\tau_{\emi}^{-1}$ for the emission process of a single TLS depends on the occupation $n_j$ of the mode $j$ actually involved. However, as the process couples the TLS to a continuum of radiation degrees of freedom, the total probablility is given by the average $\bar{n}$ of the occupations of all modes in the interval $[\Om-\D/2,\Om+\D/2]$ belonging to the state $|\Psi(t)\rangle$ as $\tau^{-1}_\emi=\g'(\bar{n}(t)+1)$ (see \cite{c-t} and section~\ref{open}). As we don't know into which mode the emission occurs, we can simplify the description of $|\Psi(t)\rangle$ given in \eqref{fullstate} as
\beq
\Psi(t)= |s_1,\ldots s_M;\bar{n}\rangle,
\label{fullstate1}
\eeq
which already is a form of coarse-graining procedure applied to the microstate \eqref{fullstate}. The next and crucial approximation is the assumption that the radiation processes of the several TLS are independent, although the coupling to the radiation field introduces statistical correlations between them. The approximation is justified by the fact that we have a {\it continuum} of radiation modes which do not interact among themselves. This implies that each individual radiation process can be considered as independent from all others; the chance of two TLS to interact with exactly the same microscopic field mode $j$ within observation time is vanishingly small. Thus we consider the joint probability $P^e_l(t,\bar{n})$ for the $l$-th TLS to be in the excited state and the average mode occupation to be $\bar{n}$. This probability satisfies the master equation \cite{reichl}
\beq
\frac{\rd P_l^e(t,\bar{n})}{\rd t}=
  -\g'P_l^e(t,\bar{n})(\bar{n} +1)
  +\g'P_l^g(t,\bar{n})\bar{n},
  \label{master}
  \eeq
  where $P_l^g(t,\bar{n})=P(\bar{n},t)-P_l^e(t,\bar{n})$ is the probablility that the $l$-th TLS is in its ground state. This relation is a consequence of the factorization property
  \beq
  P_l^e(t,\bar{n})=P^e_l(t)P(\bar{n},t),
  \label{factor}
  \eeq
which follows from the same argument as the statistical independence of the TLS, {\it i.e.} the presence of macroscopically many different modes of the radiation field in the frequency interval $[\Om-\D,\Om+\D]$, rendering the average occupation $\bar{n}$ statistically independent from the state of a single TLS. Summing \eqref{master} over $\bar{n}$, we obtain
\beq
\frac{\rd P_l^e(t)}{\rd t}=
  -\g'P_l^e(t)(\langle n\rangle(t) +1)
  +\g'(1-P_l^e(t))\langle n\rangle(t).
  \label{master2}
\eeq
Noting that $P^e_l(t)=\langle m(t)\rangle/M$, we obtain equation~\eqref{bb-m} for the ensemble average of the number $m(t)$ of excited TLS. The corresponding rate equation \eqref{bb-n} for the average photon number follows directly from the conservation of the total excitation number
  \beq
  N_{\textrm{exc}}=\sum^M_{l=1} s_l + \sum_j n_j,
  \label{exno}
  \eeq
but contains a different rate constant $\g$ to account for the continuous density of states around the resonance frequency, $\g=\g'/{\cal N}$ (see equation~\eqref{dos} in section~\ref{open}). The state vector \eqref{fullstate} satisfies a much more complicated master equation then \eqref{master2} and its stochastic evolution shows deviations of the actual value $m(t)$ from its average $\langle m(t)\rangle$ computed via \eqref{bb-n}, \eqref{bb-m}. But one may argue that a {\it typical} state will have a time evolution which deviates only slightly from its mean value for all times, $|m_{\textrm{typical}}(t)-\langle m(t)\rangle|/M\ra 0$ for $M\ra\infty$ \cite{goldstein,goldstein2}. This statement can be shown analytically for simple stochastic models like the Ehrenfest model \cite{baldovin} and also in purely deterministic toy models, e.g. the Kac ring model \cite{bricmont}. Moreover there is ample numerical evidence for such a behavior of typical states in more realistic models \cite{cerino}.

In our case, the ultimate justification of \eqref{bb-n}, \eqref{bb-m} rests on the fact that they lead to Planck's law for black-body radiation, one of the cornerstones of modern physics. It would be hard to argue that these equations are incorrect but in perfect agreement with all experiments due to some unknown cancellation of errors.

\subsection{Derivation of the rate equations \eqref{rateA} -- \eqref{rateM} for the open system}
\label{open}
We begin with the computation of the  loss rate  of photons with frequency $\om_j$  from reservoir $A$ via  channel 1 of  the open waveguide (see Fig.~\ref{Fig1}). In the coupling Hamiltonian, \eqref{hint1}, the process is of first order.
The initial state reads
\beq
|\p_\ini\rana = |\{n_A\}n_{Aj},\{n_B\},\{0_1\}0_{1k},\{0_2\},\{s\}\rana,
\label{dec-in}
\eeq
for a certain configuration of occupations $\{n_A\}$ for modes $j'$ in $A$ with $j'\neq j$. Here, $n_{Aj}$ is the occupation number of mode $j$. Similarly,
$\{n_B\}$ denotes a configuration in reservoir $B$. Further, $\{s\}$ is the configuration of excited ($s_l=e$) or ground states ($s_l=g$) of the $M$ TLS, $l=1,\ldots M$. 
The waveguide channels 1 and 2 are not occupied in $\p_\ini$.

The initial state is connected via the term $h_{jk}^*a_{1k}^\dagger a_{Aj}$ to the final state
\beq
|\p_\fin\rana = |\{n_A\}n_{Aj}-1,\{n_B\},\{0_1\}1_{1k},\{0_2\},\{s\}\rana.
\label{dec-fin}
\eeq
The $S$-matrix element reads then
\beq
S^{t_0}_{fi}=-2\pi i\del^{t_0}(E_\fin - E_\ini)\lan\p_\fin|H_{\inter}^1|\p_\ini\rana
\eeq
with $E_\fin-E_\ini = \hba(\om_k-\om_j)$. The interaction is assumed to take place over a time interval $t_0$. The corresponding regularized $\del$-function is (see \cite{c-t})
\beq
\del^{t_0}(E) = \frac{1}{\pi}\frac{\sin(Et_0/2\hba)}{E}.
\eeq
We have
\beq
\lan\p_\fin|H_{\inter}^1|\p_\ini\rana = h^*_{jk}\sqrt{n_{Aj}},
\eeq
and, according to standard reasoning \cite{c-t}, the transition rate is given by
\beq
\tau_\dec^{-1}(\om_j)=\frac{1}{t_0}\sum_k 4\pi^2|h_{kj}|^2\frac{t_0}{2\pi\hba}\del^{t_0}(E_\fin-E_\ini)
=\frac{2\pi}{\hba}|\hb|^2n_{Aj}\rho_1(\hba\om_j),
\eeq
where $\hb$ is the value of $h_{jk}$ for $\om_j=\om_k$, which is assumed to be constant in the interval $[\Om-\D/2,\Om+\D/2]$. Averaging over all modes $j$ with frequency $\om$, we obtain $\tau_\dec^{-1}(\om)=\g_\dec(\om)\na(\om)$ with
\beq
\g_\dec(\om)=\frac{2\pi}{\hba}|\hb|^2\rho_{wg}(\hba\om),
\eeq
assuming $\rho_1=\rho_2=\rho_{wg}$.
Adding the contribution of the decay into channel 2, the master equation for $\na(\om)$
describing the loss process reads
\beq
\frac{\rd \na(\om)}{\rd t}=-2\g_\dec(\om)\na(\om)+\ldots 
\label{decay}
\eeq
A similar expression holds for reservoir $B$ with the same rate constant $\g_\dec$.

The TLS couple to the channels via spontaneous emission. The corresponding first-order process leads to
\beq
\frac{\rd \langle m\rangle}{\rd t}=-(\gt_{11}(\Om)+\gt_{12}(\Om))\langle m\rangle +\ldots,
\label{decay-m}
\eeq
with $\gt_{1r}(\om)=2\pi \gb_r^2(\om)\rho_{wg}(\hba\om)/\hba$ for $r=1,2$ (see below).

We compute now the coherent transfer of a photon from reservoir $A$ to $B$  through the chiral waveguide, which is of second order in the coupling $|\hb|^2$.
  As this process concerns a   wavepacket of finite width and takes place in a finite time interval, only the right-moving channel 1 is relevant, i.e., the coupling Hamiltonian is
\beq
H^1_{\inter}=\sum_{q=A,B}\sum_{j,k} h_{jk}\left(a_{qj}^\dagger a_{1k} +\hc\right).
\eeq
We denote the initial state as
\beq
|\p_\ini\rana = |\{n_A\}n_{Aj},\{n_B\}n_{Bl},\{0_1\}0_{1k},\{0_2\},\{s\}\rana.
\eeq
The transition from $A$ to $B$ occurs via the intermediate states
\beq
|\p_\im\rana = |\{n_A\}n_{Aj}-1,\{n_B\}n_{Bl},\{0_1\}1_{1k},\{0_2\},\{s\}\rana
\eeq
towards the final state
\beq
|\p_\fin\rana = |\{n_A\}n_{Aj}-1,\{n_B\}n_{Bl}+1,\{0_1\}0_{1k},\{0_2\},\{s\}\rana,
\eeq
involving the operators
\beq
h^*_{jk}\ad_{1k}a_{Aj}, \quad h_{lk}\ad_{Bl}a_{1k}. 
\eeq
The  $S$-matrix $S_{fi}$ connecting initial and final states reads
\beq
S^{t_0}_{fi}= -2\pi i \del^{t_0}(E_\fin-E_\ini)\lim_{\eta \ra 0^+}
\sum_{k}\frac{V_{fk}V_{ki}}{E_{\ini}-E_{\im} +i\eta},
\eeq
and $E_\ini-E_\im = \hba(\om_j-\om_k)$. The matrix elements $V_{fk}$ and $V_{ki}$ are
\beq
V_{fk}=\lan\p_{\fin}|h_{lk}\ad_{Bl}a_{1k}|\p_{\im}\rana, \quad
V_{ki}=\lan\p_{\im}|h^*_{jk}\ad_{1k}a_{Aj}|\p_{\ini}\rana.
\eeq
We find
\beq
S^{t_0}_{fi} \approx -2\pi^2\del^{t_0}(\hba(\om_l-\om_j)) |\hb|^2\rho_{wg}(\hba\om_j)
\sqrt{n_{Aj}(n_{Bl}+1)},
\eeq
and obtain for the transition rate out of state $|\p_\ini\rana$,
\beq
\tau^{-1}_{jA\ra B}=\frac{1}{t_0}\sum_l|S^{t_0}_{fi}|^2=
\sum_l \frac{2\pi^3}{\hba}\del^{t_0}(\hba(\om_l-\om_j))|\hb|^4\rho^2_{wg}(\hba\om_j)n_{Aj}(n_{Bl}+1).
\eeq
We denote with $\nb(\om)$ the average occupation number per mode in reservoir $B$ at frequency $\om$. It follows that
\beq
\tau^{-1}_{jA\ra B}= \frac{2\pi^3}{\hba}|\hb|^4\rho_{wg}^2(\hba\om_j)
\rho_B(\hba\om_j)n_{Aj}(\nb(\om_j)+1).
\eeq
If we average over all modes $j$ in $A$ with frequency $\om$, the transition rate from $A$ to $B$ at $\om$ reads
\beq
\frac{1}{\tau_{A\ra B}(\om)}= \frac{2\pi^3}{\hba}|\hb|^4\rho_{wg}^2(\hba\om)
\rho_B(\hba\om)\na(\om)(\nb(\om)+1).
\eeq
Similarly, the transition from $B$ to $A$, proceeding via channel 2, is
\beq
\frac{1}{\tau_{B\ra A}(\om)}= \frac{2\pi^3}{\hba}|\hb|^4\rho_{wg}^2(\hba\om)
\rho_A(\hba\om)\nb(\om)(\na(\om)+1).
\eeq
Finally, the master equation for reservoir $A$ characterizing direct transitions between $A$ and $B$ through the waveguide is given by
\beq
\frac{\rd \na(\om)}{\rd t} = -\frac{1}{\tau_{A\ra B}(\om)} +\frac{1}{\tau_{B\ra A}(\om)} = \g_0(\om)(\nb(\om) - \na(\om)),
\label{direct}
\eeq
where we have assumed identical densities of states in $A$ and $B$, $\rho_A=\rho_B=\rho$. The rate $\g_0(\om)$ is defined as
\beq
\g_0(\om)= \frac{2\pi^3}{\hba}|\hb|^4\rho_{wg}^2(\hba\om)
\rho(\hba\om).
\eeq
The ratio of second and first order contributions follows as
\beq
\frac{\g_0(\om)}{\g_\dec(\om)}=\pi^2|\hb|^2\rho_{wg}(\hba\om)\rho(\hba\om).
\label{ratio}
\eeq

Next, we consider the absorption of radiation from reservoir $A$ by the $l$-th TLS. The initial state is
\beq
|\p_\ini\rana = |\{n_A\}n_{Aj},\{n_B\},\{0_1\}0_{1k},\{0_2\},\{s\}g_l\rana.
\label{absorption}
\eeq
The absorption of a wavepacket of finite spatial extension can only proceed via channel 1. The intermediate states read then
\beq
|\p_\im\rana = |\{n_A\}n_{Aj}-1,\{n_B\},\{0_1\}1_{1k},\{0_2\},\{s\}g_l\rana
\eeq
and the final state is obtained by absorption of the photon in channel 1 by the TLS,
\beq
|\p_\fin\rana = |\{n_A\}n_{Aj}-1,\{n_B\},\{0_1\}0_{1k},\{0_2\},\{s\}e_l\rana.
\eeq
For the S-matrix element we have
\beq
S^{t_0}_{fi}=-2\pi^2\hb \gb_1\del^{t_0}(\hba\Om-\hba\om_j)\rho_{wg}(\hba\om_j)
\sqrt{n_{Aj}},
\eeq
where $\gb_1$ is the value of $g_{1k}$ for $\om_k=\om_j$. The corresponding term in the master equation for $\langle m\rangle$, the average number of excited TLS, is obtained by summing over all initial states, which leads to the expression
\beq
\frac{\rd \langle m\rangle}{\rd t} = -\frac{2\pi^3}{\hba}(M-\langle m\rangle)\hb^2\gb^2_1\rho_{wg}^2(\hba\Om)\rho(\hba\Om)\na(\Om) +\ldots
\eeq
In an anlogous manner, the radiation from reservoir $B$ is absorbed via channel 2 by the TLS,
\beq
\frac{\rd \langle m\rangle}{\rd t} = -\frac{2\pi^3}{\hba}(M-\langle m\rangle)\hb^2\gb^2_2\rho_{wg}^2(\hba\Om)\rho(\hba\Om)\nb(\Om) +\ldots
\eeq
On the other hand, the emission from the TLS towards reservoir $A$ must proceed via the left-moving channel 2. The initial state is now
\beq
|\p_\ini\rana = |\{n_A\}n_{Aj},\{n_B\},\{0_1\},\{0_2\}0_{2k},\{s\}e_l\rana,
\eeq
the intermediate state
\beq
|\p_\im\rana = |\{n_A\}n_{Aj},\{n_B\},\{0_1\},\{0_2\}1_{2k},\{s\}g_l\rana,
\eeq
and the final state
\beq
|\p_\fin\rana = |\{n_A\}n_{Aj}+1,\{n_B\},\{0_1\},\{0_2\}0_{2k},\{s\}g_l\rana.
\eeq
A calculation completely analogous to the one for absorption above leads to the following term in the master equation for $\langle m\rangle$, this time summing over final modes in reservoir $A$,
\beq
\frac{\rd \langle m\rangle}{\rd t} = \frac{2\pi^3}{\hba}\langle m\rangle\hb^2\gb^2_2\rho_{wg}^2(\hba\Om)\rho(\hba\Om)(\na(\Om)+1) +\ldots
\label{stim-em}
\eeq
In this expression, the  term proportional to $\na$ is associated with stimulated emission from the TLS. It is noteworthy that the stimulated emission of the TLS is not caused by photons that have been emitted by $A$ and then impinge on the TLS to create photons in the same mode. Here, in contrast, the stimulated emission of the TLS is induced by photons that are \textit{received} by $A$ after having been emitted by the TLS.
This counterintuitive effect, which is solely due to the Bose statistics and therefore only possible in quantum physics, restores detailed balance for the case of a cavity embedded in another one, in accord with Kirchhoff's law on black body radiation (see section \ref{embed}).

With the definition
\beq
\gt_r(\Om)=\frac{2\pi^3}{\hba}\hb^2\gb^2_r\rho_{wg}^2(\hba\Om)\rho(\hba\Om)
\eeq
for $r=1,2$ we obtain the master equation for $\langle m\rangle$,
\begin{align}
\frac{\rd \langle m\rangle}{\rd t}&= -[\gt_{11}(\Om) +\gt_{12}(\Om)]\langle m\rangle + \gt_1(\Om)\left[\na(M-\langle m\rangle)-(\nb+1)\langle m\rangle\right] \nn\\ 
&+\gt_2(\Om)\left[\nb(M-\langle m\rangle)-(\na+1)\langle m\rangle\right],
\label{rate-me}
\end{align}
which is Eq.\eqref{rateM} in the main text. To compute the terms corresponding to the absorption and emission processes in the rate equations for $\na$ and $\nb$, we note that the rates $\gt_r$ contain a summation over initial, respectively final modes in the reservoirs. The coefficients $\g_r$ describing the temporal change in the average occupation number per mode, $\lana n_q\rana$, are therefore $\gt_r$ divided by the number of modes in the relevant frequency interval. Thus,
\beq
\N = \hba\int_{\Om-\D/2}^{\Om+\D/2}\rd\om \rho(\hba\om), \quad \g_r=\gt_r/\N.
\label{dos}
\eeq
Together with \eqref{decay} and \eqref{direct}, it follows for $\om=\Om$,
  \begin{align}
    \frac{\rd \na}{\rd t} &= -2\g_\dec\na + \g_0(-\na+\nb)- \g_1(M-\langle m\rangle)\na\nn\\
    &+ \g_2\langle m\rangle(\na+1),\label{rateAa}\\
    \frac{\rd \nb}{\rd t} &= -2\g_\dec\nb +  \g_0(-\nb+\na)- \g_2(M-\langle m\rangle)\nb\nn\\
    &+ \g_1\langle m\rangle(\nb+1), \label{rateBa}
\end{align}
  which are Eqs. \eqref{rateA} and \eqref{rateB}.

  \subsection{Derivation of the entropy production \eqref{sab} -- \eqref{sbt}}
  \label{entr-prod}
  As no work is done during the irreversible process, we can write for the heat flow out of the system within our approximation
  \beq
  Q^{\ext}=-\frac{\rd}{\rd t}\big(\lana H_A\rana+\lana H_B\rana+\lana H_{\tls}\rana\big) >0,
  \eeq
  because the coherences of the TLS are not important and the waveguide contains no photons on average.
  We consider in the following only quantities (entropy and energy) corresponding to the interval
  $[\hba(\Om-\D/2),\hba(\Om+\D/2)]$ because other photon modes do not interact with the TLS. The heat transferred to the environment per unit time reads for each subsystem (see \eqref{rateA} -- \eqref{rateM})
  \beq
  Q^{\ext}_q=2\hba\Om{\N}\g_{\dec}\lana n_q\rana, \quad
  Q^{\ext}_{\tls}=\hba\Om(\gt_{11}+\gt_{12})\langle m\rangle.
  \eeq
  The corresponding entropy change of the system is given by
  \beq
  -\int\rd{\bm o}\cdot{\bm J}_{\sys} = -\left(\sum_{q=A,B}\frac{Q^{\ext}_q}{T_q} + \frac{Q^{\ext}_{\tls}}{T_{\tls}}\right),
  \eeq
  with
  \beq
  T_q^{-1}=\frac{k_B}{\hba\Om}\ln\left(\frac{\lana n_q\rana +1}{\lana n_q\rana}\right), \quad
  T_{\tls}^{-1}=\frac{k_B}{\hba\Om}\ln\left(\frac{M-\langle m\rangle}{\langle m\rangle}\right).
  \eeq
  The entropy of the system is always reduced due to the vanishing temperature of the bath. Although the entropy production of system plus bath is thus formally positive and infinite, the actual entropy loss of the system stays finite and is of course compatible with the second law which mandates that the local entropy production within the system must be non-negative. In our case, the local entropy production is induced by the heat exchange between the three subsystems. The heat change of reservoir $A$ due to the interaction with reservoir $B$ reads
  \beq
  Q^A_{A,B}=-Q^B_{A,B}=\hba\Om\N\g_0(\nb-\na).
  \eeq
  It follows for the entropy production due to this process
  \beq
  \s_{A,B}=\hba\Om\N\g_0(\nb-\na)(T_A^{-1}-T_B^{-1}),
  \eeq
  which is equation \eqref{sab}. The equations \eqref{sat}, \eqref{sbt} for the heat exchange between the reservoirs and the TLS are deduced correspondingly.
Alternatively, one may compute the entropy change in each subsystem directly with the expressions \eqref{entropyTLS} for the TLS entropy and \eqref{entropy2} for the entropy of the radiation modes. This shows that the entropy change of each subsystem is not accompanied by additional entropy production but is solely due to heat transfer between subsystems, a consequence of the local thermal equilibrium in  $A$, $B$ and the TLS.
  
\subsection{Derivation of the rate equations for the closed system}
\label{closed}
\begin{figure}
	\centering
	\includegraphics[width=0.4\textwidth]{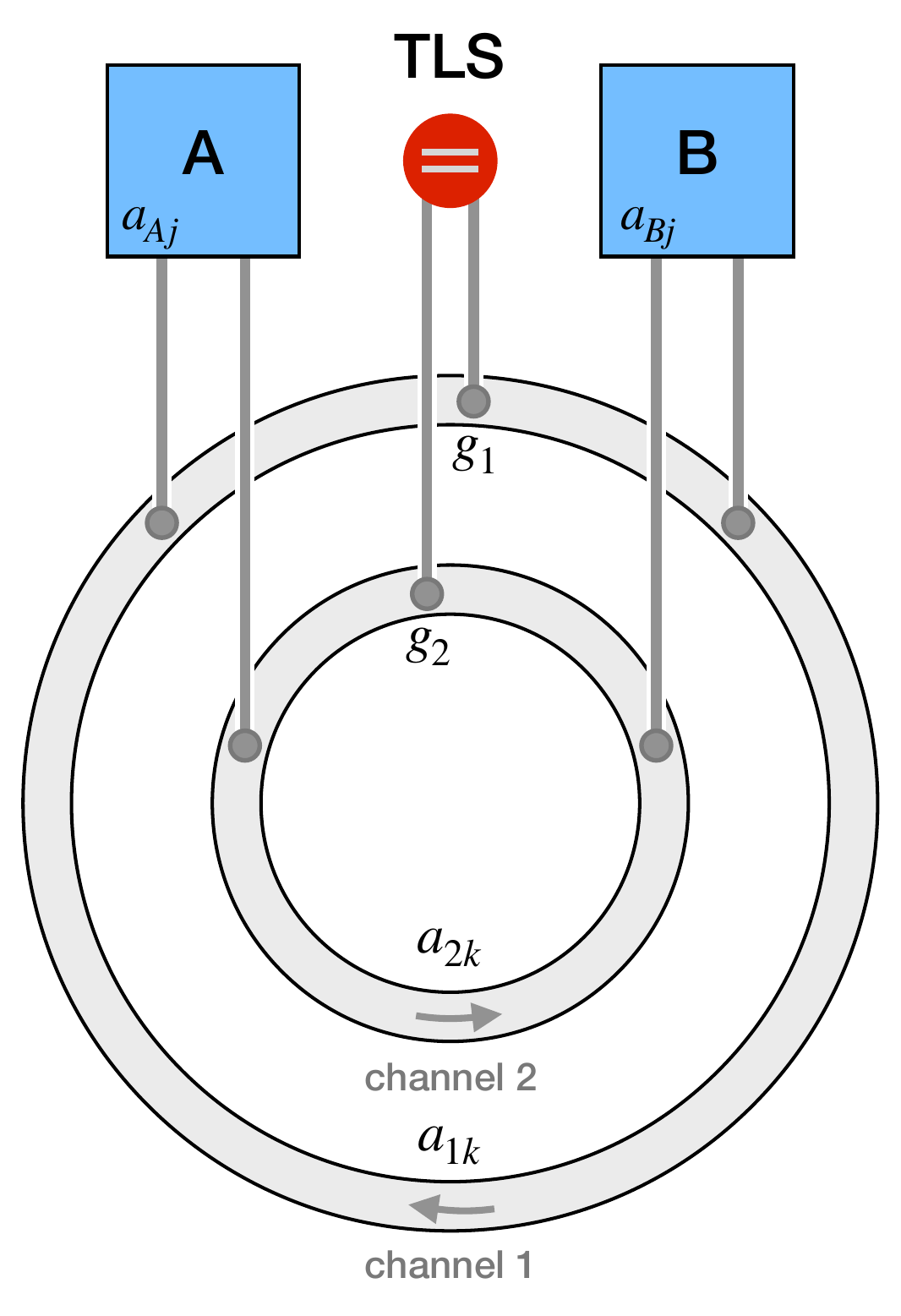}
	\caption{The system obtained by closing the open waveguide of the system shown in Fig.~1 of the main text. The occupation numbers of the chiral channels 1 and 2 do not vanish due to equilibration with the reservoirs $A$ and $B$.}
	\label{loop}
\end{figure}

This chapter discusses a variant of the open system characterized
  in section~\ref{open}. In this variant, which is a closed system, the
  waveguide satisfies periodic boundary conditions, corresponding to a  loop
  of length $L$, where $L$ is much larger than the distance between the
  reservoirs $A$ and $B$, see Fig.~\ref{loop}. We show that a description
  assuming real absorption and emission processes fulfills the detailed
  balance condition to first order in the coupling. Second order processes
  analogous to those described in Eqs. \eqref{absorption} -- \eqref{stim-em} break the detailed balance condition for the case that coherent absorption and emission is only possible along the short path between the reservoirs and the TLS, while dephasing occurs for  wave packets emitted by a TLS and traveling the long way around the loop before reaching one of the reservoirs. The latter case is  already accounted for by the first-order terms describing the equilibration between the reservoirs/TLS and the waveguide. Of course, if the dynamics of the closed system is considered to be unitary, corresponding to completely coherent evolution, the entropy does not change. This situation could be approximated by treating all second order processes as coherent, 
including those on the long path between the TLS and the cavities. Then 
the detailed balance condition would be satisfied, leading to stabilization of the state with maximal entropy. However, if the coherence is restricted to processes occurring along the short path, the ensuing steady state does not have maximum entropy.
In the closed system, the occupancy of the waveguide can no longer be assumed to be zero, as the photons cannot escape towards infinity.
We describe the occupancy in channel $q$ by $\lana n_q\rana$ for $q=1,2$.
The Hamiltonian is given in Eqs. \eqref{ham} -- \eqref{hint2}.
The exchange of photons between cavity $A$ and channel 1 of the waveguide  
proceeds via a first-order process analogous to that given in Eqs. \eqref{dec-in} and \eqref{dec-fin}, but the matrix element reads now (suppressing the frequency arguments),
\beq
\lan\p_\fin|H_{\inter}^1|\p_\ini\rana = h^*_{jk}\sqrt{n_{Aj}(n_1+1)}.
\eeq
This gives for the transition rate from mode $j$ in $A$ to channel 1 of the waveguide
\beq
\tau_{A\ra 1}^{-1}=\g_{\dec}n_{Aj}(\nei+1),
\eeq
and for the reverse process,
\beq
\tau_{1\ra A}^{-1}=\g_{\dec}\nei(n_{Aj} +1),
\eeq
where in this case a sum over initial states has to be performed to obtain the rate of emission into the fixed mode $j$ of $A$.
The terms in the master equation for $\na$ are thus
\beq
 \frac{\rd \na}{\rd t} = \g_{\dec}(\nei -\na) + \ldots, 
\eeq
and for $\nei$
\beq
 \frac{\rd \nei}{\rd t} = \g_{\dec}'(\na -\nei) + \ldots, 
 \label{channel1}
\eeq
with the rate constant
$\g_\dec'=(\rho/\rho_{wg})\g_\dec$. Analogous expressions are obtained for $B$ and channel 2.
Another first-order process couples the TLS and the waveguide modes. We find for this contribution to the rate equation for $\langle m\rangle$
\beq
 \frac{\rd \langle m\rangle}{\rd t} = \sum_{r=1,2} \gt_{1r}[(M-\langle m\rangle)\nr -\langle m\rangle(\nr +1)] + \ldots,
 \eeq
 with $\gt_{1r}=2\pi g_r^2\rho_{wg}/\hba$ (compare Eq.\eqref{decay-m}).
 The corresponding terms in the rate equations for $\nr$ are
 \beq
  \frac{\rd \nr}{\rd t} = \g_{1r}[\langle m\rangle(\nr +1) -(M-\langle m\rangle)\nr] + \ldots,
\eeq
and
\beq
\g_{1r}=\gt_{1r}/\N', \quad \N' = \hba\int_{\Om-\D/2}^{\Om+\D/2}\rd\om \rho_{wg}(\hba\om).
\eeq
All these first-order terms satisfy the detailed balance condition. They lead naturally to thermal equilibration  between the reservoirs and the waveguide.
For the second-order terms, we have first the process described by  Eq.\eqref{rate-me},  
\begin{align}
\frac{\rd \langle m\rangle}{\rd t}& = \gt_1(\nei+1)\left[\na(M-\langle m\rangle)-(\nb+1)\langle m\rangle\right] \nn\\
 &+\gt_2(\nz+1)\left[\nb(M-\langle m\rangle)-(\na+1)\langle m\rangle\right] + \ldots,
\label{rate-mec}
\end{align}
which depends  also on the occupation numbers $\nr$ of the waveguide. This term is accompanied by corresponding terms in the rate equations for the reservoirs. It does not satisfy detailed balance because we have only considered the short path between the reservoirs and the TLS (the only available one in the open system). Including also the long path around the circular waveguide would again reinstate the detailed balance condition. We assume  that this second process is not coherent due to dephasing of the photon while traveling along the loop. 
Such a dephasing may, for example, be caused by scattering processes induced in the long section of the loop.
The second order term given in Eq.\eqref{direct} is modified in the closed system as follows,
\beq
\frac{\rd \na}{\rd t}=\g_0\big( (\nz+1)(\na+1)\nb -(\nei+1)(\nb+1)\na\big) +\ldots,
\eeq
together with an equivalent term for $\nb$.
Another term of second order in $|\hb|^2$ couples the two channels of the waveguide via a reservoir.
For channel 1 it reads
\beq
\frac{\rd \nei}{\rd t} = \g_3(\na+\nb+2)(\nz-\nei) +\ldots,
\eeq
with $\g_3=2\pi^3|\hb|^4\rho^2\rho_{wg}/\hba$.
Finally, there is a  term connecting the channels via an intermediate excitation of the TLS, proportional to $g_1^2g_2^2$. This term can be neglected if one of the chiral couplings $g_r$ is close to zero, as we have assumed in the numerical evaluation.
Collecting all the terms, we obtain the rate equations for the photon occupation numbers,
\begin{align}
\frac{\rd \na}{\rd t}&=\g_{\dec}(\nei+\nz-2\na) -\g_1(M-\langle m\rangle)\na(\nei+1)\nn\\
 & +\g_2\langle m\rangle(\na+1)(\nz+1) +\g_0\big( (\nz+1)(\na+1)\nb \nn\\ 
 &-(\nei+1)(\nb+1)\na\big), \label{rateAc}\\
 \frac{\rd \nb}{\rd t}&=\g_{\dec}(\nei+\nz-2\nb) -\g_2(M-\langle m\rangle)\nb(\nz+1)\nn\\
 & +\g_1\langle m\rangle(\nb+1)(\nei+1) +\g_0\big( (\nei+1)(\nb+1)\na \nn\\ 
 & -(\nz+1)(\na+1)\nb\big), \label{rateBc}\\
 \frac{\rd \nei}{\rd t}&=\g_\dec'(\na+\nb-2\nei)-\g_{11}\big((M-\langle m\rangle)\nei
 - \langle m\rangle(\nei+1)\big) \nn\\
  &+\g_3(\na+\nb+2)(\nz - \nei),\label{rateE} \\
 \frac{\rd \nz}{\rd t}&=\g_\dec'(\na+\nb-2\nz)-\g_{12}\big((M-\langle m\rangle)\nz
 - \langle m\rangle(\nz+1)\big)  \nn\\
  &+\g_3(\na+\nb+2)(\nei - \nz). \label{rateZ}
\end{align}
The average number of excited TLS is determined by
\begin{align}
 \frac{\rd \langle m\rangle}{\rd t}& = \sum_{r=1,2} \gt_{1r}[(M-\langle m\rangle)\nr -\langle m\rangle(\nr +1)]\nn\\
 &-\langle m\rangle\big(\gt_1(\nb+1)(\nei+1)+\gt_2(\na+1)(\nz+1)\big) \label{rateMc}\\
 &+(M-\langle m\rangle)\big(\gt_1\na(\nei+1)+\gt_2\nb(\nz+1)\big).\nn
 \end{align}
\begin{figure}
	\centering
	\includegraphics[width=0.7\textwidth]{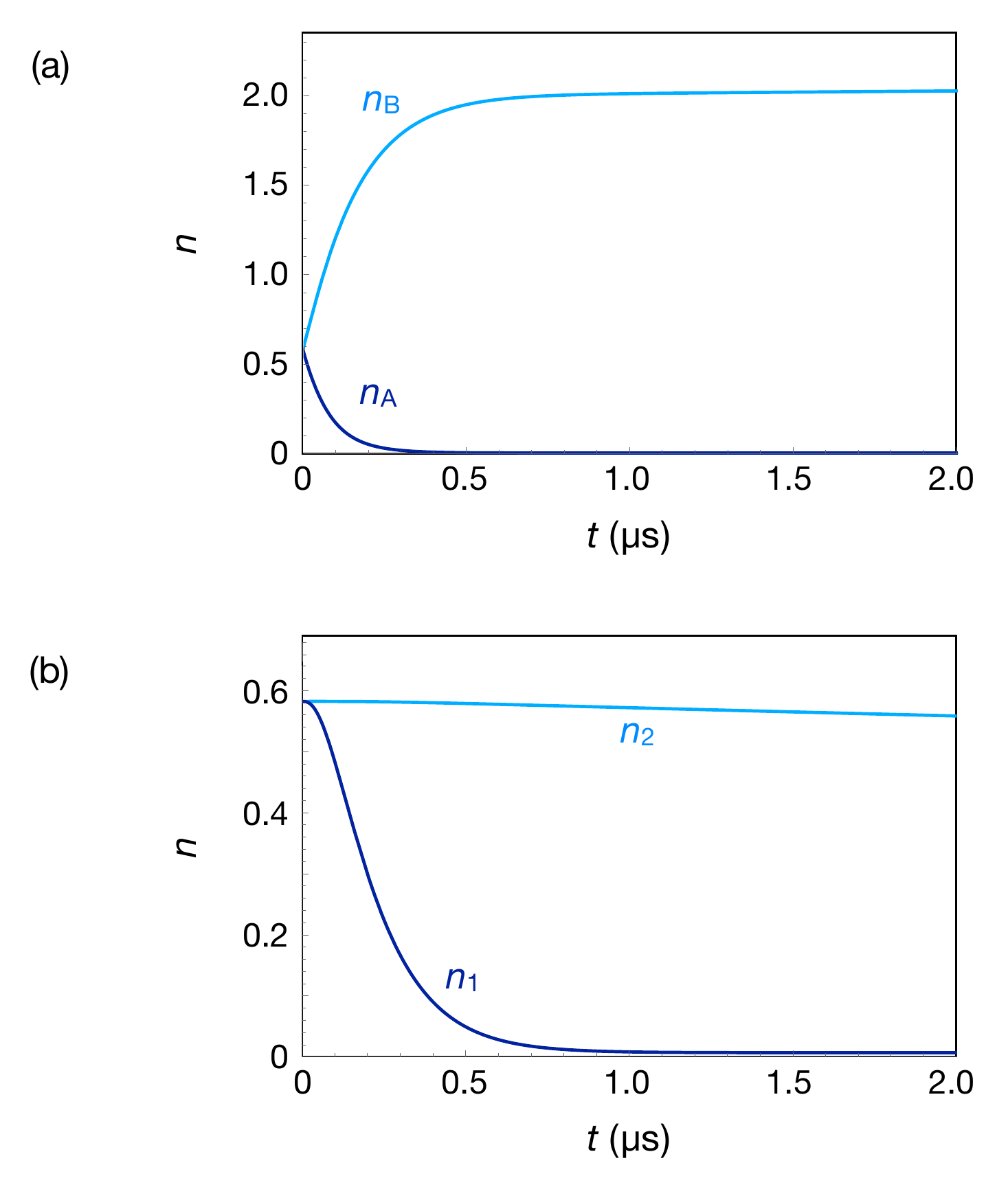}
	\caption{Solutions of the rate equations \eqref{rateAc}-\eqref{rateMc} as function of time, starting from initial thermal equilibrium. Panel (a) displays $\na(t)$, $\nb(t)$ and panel (b) $\nei(t)$, $\nz(t)$. 
The average occupations $\lana n_q\rana(t)$ per mode reach a novel steady state with $\na\neq\nb$. The photon densities in reservoir $A$ and channel 1 fall to zero, whereas channel 2 stays occupied and reservoir $B$ is populated. The displayed time interval corresponds to the time scale set by the coupling to the TLS. Parameters used are
	$\g_\dec=\g_\dec'=\g_0=\g_3=10$\;kHz, $\gt_1=\gt_{11}=10$\;MHz, $\g_1=\g_{11}=100$\;kHz, $\gt_2=\gt_{12}=\g_2=\g_{12}=0$ and $\N=\N'=100$, $\hba\Om/k_BT(0)=1$.}
	\label{occup-short}
\end{figure}
To compute the total entropy of the system, we note first that
\beq
S_M=-k_BM\big(p_e\ln p_e +(1-p_e)\ln (1-p_e)\big),
\label{entropyTLSa}
\eeq
is the entropy of the TLS, as given by Eq.\eqref{entropyTLS}.
The entropy $S_\rad(\om)$ of the radiation per mode in the reservoirs and the waveguide depends only on $\lana n_q(\om)\rana$ and reads
\beq
S^l_{\rad}(\om) =\frac{\hba\om}{T}\langle n_l(\om)\rana +k_B\ln(1+\langle n_l(\om)\rana),
\label{entropy2}
\eeq
for $l=A,B,1,2$. Because the temperature depends on $\lana n_l\rana$ as described by Eq.\eqref{n-T} in the main text, the entropy is only a function  of $\lana n_l\rana$. The effective entropies and temperatures in the reservoirs and the waveguide are computed  under the assumption that the photons in each reservoir/channel  thermalize in the usual way quickly as an ideal Bose gas.
We approximate the total entropy by the sum of the entropies of the subsystems, as is justified for large $\lana n_q\rana$ and $M$.
The total entropy is then given by
 \beq
S(t)=S_M(t) + \int_{\Om-\D/2}^{\Om+\D/2}\rd\om [\rho(\om)S^R(\om,t)+\rho_{wg}(\om)S^{wg}(\om,t)],
\label{entropy3}
\eeq
with
\beq
S^R=S^A_\rad+S^B_\rad, \quad S^{wg}=S^1_\rad +S^2_\rad.
\eeq
For constant densities of states $\rho$ and $\rho_{wg}$, we may write
\beq
S(t) =S_M(t) + \N S^R(\Om,t) + \N' S^{wg}(\Om,t).
\eeq
The temporal evolution of the closed system, starting with thermal equilibrium, is depicted in  Fig.~\ref{occup-short} for the same parameters as in the open system. The non-reciprocal interaction with the TLS empties reservoir $A$  and the active (coupled) channel 1, while the occupation of reservoir $B$ rises. The inert channel 2 is unaffected on this short timescales. It interacts via the weak couplings $\g_\dec'$ and $\g_3$ with $B$, which manifests only on much longer timescales, as depicted in Fig.~\ref{occup-long}. This separation of timescales has the same origin in the closed and the open system.
\begin{figure}
	\centering
	\includegraphics[width=0.7\textwidth]{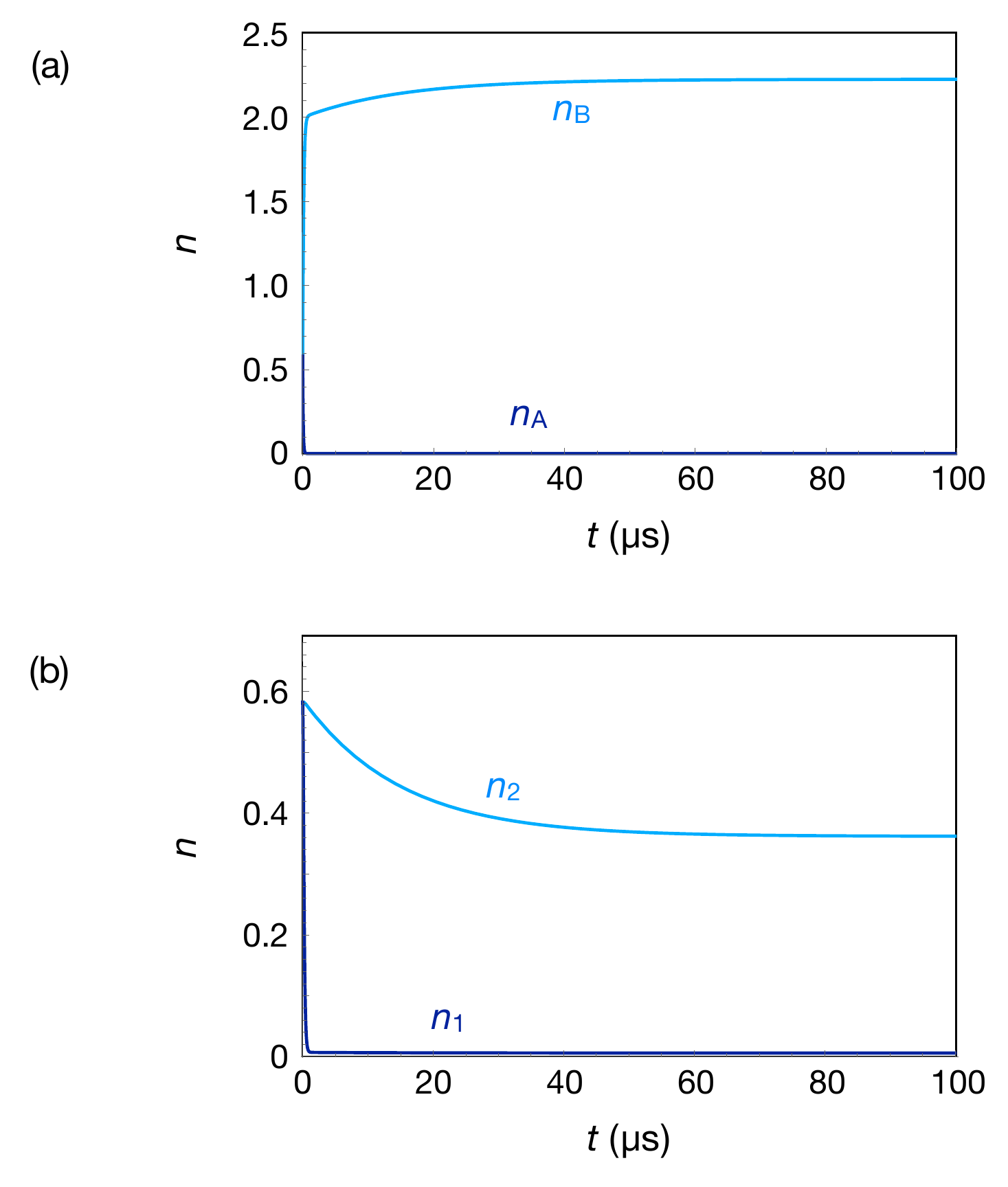}
	\caption{The asymptotic temporal behavior of reservoir $B$ (panel (a)) and the inert channel 2 (panel (b)).
	The occupation in $A$ and channel 1 is almost zero.  The novel steady state has unequal occupations in all photonic subsystems and therefore lower entropy than the initial state. The parameters used are the ones of Fig.~\ref{occup-short}.}
	\label{occup-long}
\end{figure}

 \subsection{Detailed balance for a system with an embedded cavity}
 \label{embed}
\begin{figure}
	\centering
	\includegraphics[width=0.7\textwidth]{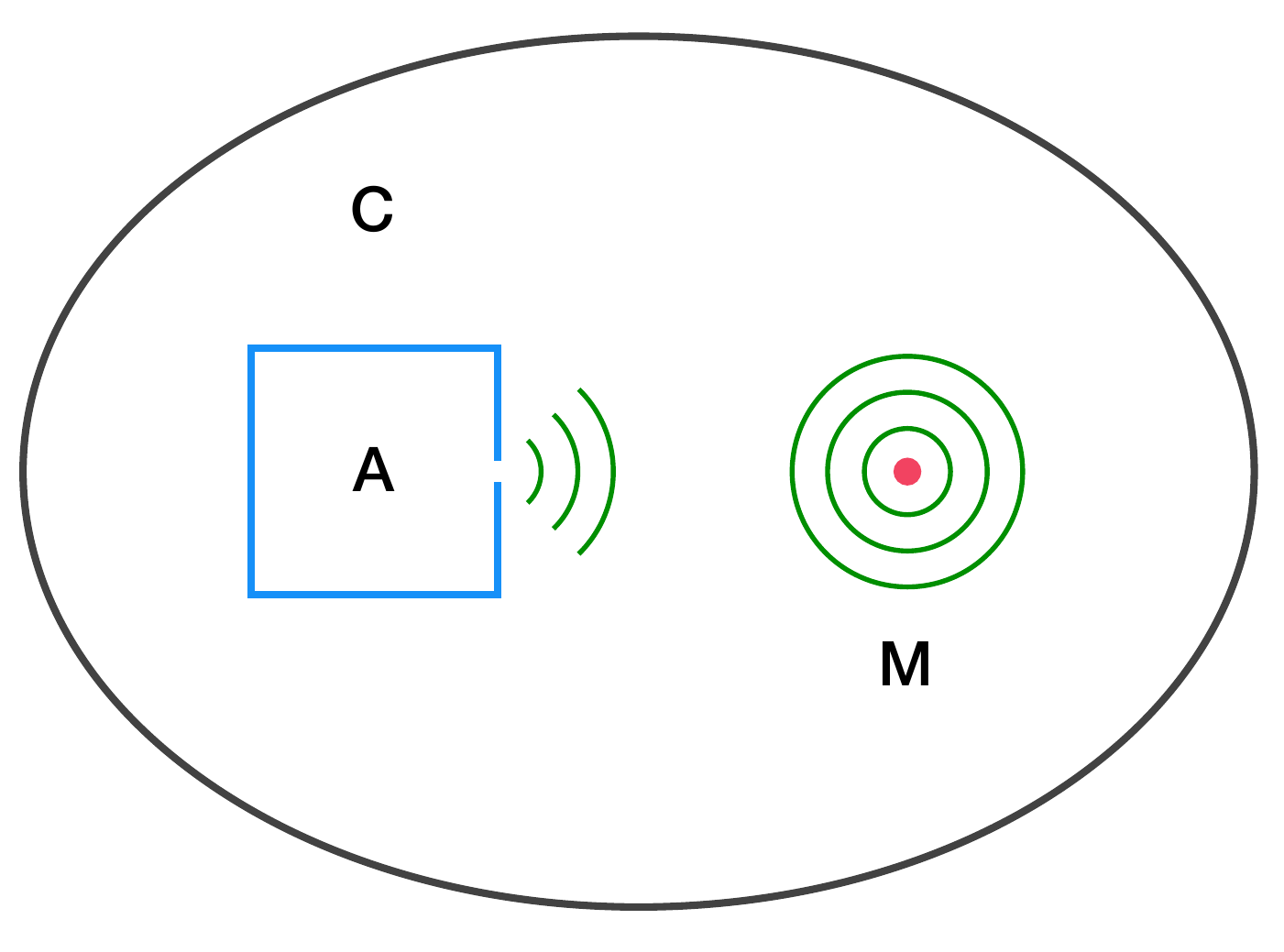}
	\caption{The cavity system. The small cavity $A$ exchanges radiation with the surrounding closed cavity $C$. The collection $M$ of two-level systems interacts with the radiation modes of $C$.}
	\label{supp-fig}
\end{figure}

We shall now demonstrate that the term in \eqref{stim-em} corresponding to radiation stimulated by the receiving reservoir leads to the detailed balance condition in case a cavity is embedded into another one. Here, detailed balance is also obtained in the second-order terms, in contrast to the chiral system treated in the previous section. We consider a closed cavity $C$ with adiabatic walls. Inside of $C$ there is a smaller cavity $A$ which is coupled to $C$ through a small opening. Besides $A$, a collection of $M$ two-level systems is located in $C$ (Fig.~\ref{supp-fig}).
 The Hamiltonian of this system is given as
\beq
H=H_A + H_C + H_{\tls} + H^1_{\inter} +  H^2_{\inter},
\label{hams}
\eeq
where
\beq
H_q=\hba\sum_{j}\om_{qj} a_{qj}^\dagger a_{qj}, \qquad
H_{\tls}=\frac{\hba\Omega}{2}\sum_{l=1}^M\s^z_l
\eeq
for $q=A,C$.
The interaction between $A$ and $C$ is given by
\beq
H^1_{\inter}=\sum_{j,k}h_{jk}a^\dagger_{Aj}a_{Ck} + \hc,
\label{hint1s}
\eeq
and $C$ interacts with $M$ as
\beq
H^2_{\inter}=\sum_{l=1}^M \sum_{k} g_{k}a_{Ck}\s^+_l + \hc
\label{hint2s}
\eeq
 The rate equations are computed as above, but now the exchange between $A$ and $C$ is given by terms of first order in the coupling $h_{jk}$,
 \begin{align}
    \frac{\rd \na}{\rd t} &= \g_4\left[-\na(\nc+1) + \nc(\na+1)\right] +\ldots,
\label{rateA-ei}\\
\frac{\rd \nc}{\rd t} &= \g_4'\left[-\nc(\na+1) + \na(\nc+1)\right] +\ldots
\label{rateC-ei}
\end{align}
with $\g_4'=(\rho_A/\rho_C)\g_4$, (compare Eq.\eqref{channel1}).
The interaction between the TLS and $C$ leads to the terms
 \beq
 \frac{\rd \nc}{\rd t} = -\g_5\nc(M-\langle m\rangle) +\g_5(\nc+1)\langle m\rangle +\ldots,
 \eeq
 which are of first order in $g_{k}$ and correspond to standard black-body radiation.
 Besides these first-order terms, there are also terms of second order in the couplings $h_{jk}$ and $g_k$, describing the interaction of the small cavity $A$ with the TLS via intermediate states belonging to $C$. However, in contrast to the second-order terms discussed in section \ref{closed}, these second-order terms are compatible with detailed balance.
 The corresponding terms in the rate equation for $A$ read
 \beq
 \frac{\rd \na}{\rd t} = \g_6(\nc+1)^2\left[-\na(M-\langle m\rangle) +(\na+1)\langle m\rangle\right]+\ldots
 \eeq
 As above, the term for stimulated emission, $\g_6(\nc+1)^2\na \langle m\rangle$, is not related to radiation emerging from cavity $A$ into $C$ which would have the wrong direction (see Fig.~\ref{supp-fig}), but comes from the occupation of $A$-modes in the final state. The rate equations for $A$ and $C$ are therefore
 \begin{align}
   \frac{\rd \na}{\rd t} &= \g_4\left[-\na + \nc\right] +
   \g_6(\nc+1)^2\left[-\na(M-\langle m\rangle) +(\na+1)\langle m\rangle\right],
\label{rateA-ei1}\\
\frac{\rd \nc}{\rd t} &= \g_4'\left[-\nc + \na\right] 
 -\g_5\nc(M-\langle m\rangle) +\g_5(\nc+1)\langle m\rangle.
\label{rateC-ei1}
\end{align}
 These equations fulfill the detailed balance condition and lead to thermal equilibrium between $A$, $C$ and $M$. The result is consistent with Kirchhoff's law, which states that the interior of a thermally isolated hohlraum has no influence on the final steady state of the contained radiation. This radiation  exhibits the black-body spectrum found by M. Planck.

\bibliographystyle{spphys}       
\bibliography{fp1.bib}   

\end{document}